\documentclass[12pt]{article}

\usepackage{cancel,slashed}
\usepackage{bbm}
\usepackage{amssymb}
\usepackage{amsfonts}
\usepackage{amsmath}
\usepackage{graphicx}
\usepackage{cite}

\usepackage{latexsym}
\usepackage{color}
\usepackage{xypic}
\usepackage{cancel,slashed}
\usepackage[linktocpage]{hyperref}
\usepackage{color}
\usepackage{transparent}

\newcommand{\bmat}{\left(\begin{array}}
\newcommand{\emat}{\end{array}\right)}

\def\p{\partial}
\def\a{\alpha}

\def\b{\beta}
\def\g{\gamma}
\def\d{\delta}
\def\th{\theta}

\def\vphi{\varphi}
\def\-{\hphantom{-}}

\def\s2{\frac{1}{\sqrt2}}

\def\oh{\frac{1}{2}}
\def\beq{\begin{equation}}
\def\eeq{\end{equation}}
\def\beqa{\begin{eqnarray}}
\def\eeqa{\end{eqnarray}}

\def\tr{{\rm tr \,}}

\def\ca{{\mathcal A}}

\def\cn{{\mathcal N}}

\def\Dsl{\,\raise.15ex\hbox{/}\mkern-13.5mu D} 

\def\CR {{\cal R}}

\def\CA {{\cal A}}

\def\CO {{\cal O}}

\def\tr{\mbox{Tr}}
\def\str{\mbox{STr}}

\def\be{\begin{equation}}
\def\ee{\end{equation}}
\def\bea{\begin{eqnarray}}
\def\eea{\end{eqnarray}}
\def\raw{\rightarrow}

\def\IC{\mathbb{C}}
\def\IN{\mathbb{N}}

\def\oh{\frac{1}{2}}
\def\a{{\alpha}}
\def\b{{\beta}}
\def\d{{\delta}}
\def\eps{{\epsilon}}
\def\th{{\theta}}

\def\lam{{\lambda}}

\def\sig{{\sigma}}

\def\g{{\gamma}}
\def\G{{\Gamma}}
\def\vphi{{\varphi}}
\def\p{{\partial}}

\def\vec#1{{\overrightarrow{#1}}}
\def\str{\mbox{STr}}

\def\w{{\wedge}}
\def\ph{{\partial_{\langle A\rangle}}}
\def\pa{{\bar \partial_{\langle A\rangle}}}

\def\lp{{\langle\Phi\rangle}}

\def\sm2{{\mbox{\small 2}}}

\begin{document}
\pagestyle{plain}

\makeatletter
\@addtoreset{equation}{section}
\makeatother
\renewcommand{\theequation}{\thesection.\arabic{equation}}
\pagestyle{empty}
\rightline{ IFT-UAM/CSIC-13-085}
\vspace{0.5cm}
\begin{center}
\LARGE{{Up-type quark masses in SU(5) F-theory models}
\\[10mm]}
\large{A. Font,$^1$ F. Marchesano,$^2$ D. Regalado$^{2,3}$ and G. Zoccarato$^{2,3}$ \\[10mm]}
\small{
${}^1$  Departamento de F\'{\i}sica, Centro de F\'{\i}sica Te\'orica y Computacional \\[-0.3em]
 Facultad de Ciencias, Universidad Central de Venezuela\\[-0.3em]
 A.P. 20513, Caracas 1020-A, Venezuela  \\[2mm] 
${}^2$ Instituto de F\'{\i}sica Te\'orica UAM-CSIC, Cantoblanco, 28049 Madrid, Spain \\[2mm] 
${}^3$ Departamento de F\'{\i}sica Te\'orica, 
Universidad Aut\'onoma de Madrid, 
28049 Madrid, Spain
\\[8mm]} 
\small{\bf Abstract} \\[5mm]
\end{center}
\begin{center}
\begin{minipage}[h]{15.0cm} 

F-theory SU(5) unification has been proposed as a scenario where the mass of the top quark is naturally large, as opposed to type II SU(5) models. We analyze this claim from the viewpoint of local SU(5) F-theory models, by explicitly computing the $10 \times 10 \times 5$ Yukawa couplings that are developed in the vicinity of an $E_6$ singularity. Realizing this singularity via T-branes allows for a non-trivial mass for the top quark, while lighter generations of up-type quarks still have vanishing Yukawa couplings. Nevertheless, we show that by taking instanton effects into account non-vanishing Yukawas are induced for all U-quark families, together with a hierarchical structure at the level of the superpotential. Finally, by solving for internal wavefunction profiles we compute physical U-quark Yukawa couplings and show that this F-theory scenario allows to describe the measured top quark mass, as well as the observed quotients of U-quark masses.

\end{minipage}
\end{center}
\newpage
\setcounter{page}{1}
\pagestyle{plain}
\renewcommand{\thefootnote}{\arabic{footnote}}
\setcounter{footnote}{0}


\tableofcontents


\section{Introduction}
\label{s:intro}

Given the vast set of string theory vacua one may wonder what is the appropriate strategy to draw general lessons out of realistic and semi-realistic string constructions \cite{thebook}. One would expect in particular that reproducing the Standard Model (SM) as a low energy limit of string theory provides a rationale for the disparity of couplings that define its flavor structure. In this sense it has been realized that a full knowledge of the string landscape may not be necessary to address this point. Indeed, due to the localization properties of branes certain quantities like gauge and Yukawa couplings do not depend on the full geometry of the compactification, but instead on the local data in the region where SM fields are localized. This important feature allows to implement a bottom-up approach to reproduce the SM within string theory \cite{aiqu}, in which one first specifies the local geometry that describes the SM sector and then considers all its possible global completions. Because it is in this second step that the landscape arises, one may still hope to infer a general scheme that describes the SM flavor structure from the analysis of local string theory models. 

A particular context in which this bottom-up approach can be implemented is in type IIB compactifications with D3 and/or D7-branes. A rather attractive feature of these local constructions is that, because all gauge interactions arise from the same region of the compactification space, all SM gauge couplings typically depend on the same closed string modulus and one is led naturally to a gauge coupling unification scheme. This already suggests that a promising avenue to realize the SM coupling structure in string theory is via constructing local GUT models, and in particular SU(5) GUT's whose chiral spectrum can be easily realized via D-branes.
This type IIB framework has however a serious drawback when realizing SU(5) GUT's, namely that the U(1) selection rules that are common in type II models forbid the presence of the up-like $\mathbf{10} \times \mathbf{10} \times \mathbf{5}$ Yukawa coupling at the perturbative level \cite{thebook}. While one may still generate this coupling via D-brane instantons \cite{bclrt08}, the large experimental value for the top Yukawa hints at an scenario where up-like Yukawa couplings are generated on equal footing as down-like Yukawas. 

In this respect local F-theory SU(5) GUT models have emerged as a very promising scenario \cite{bhv1,bhv2,dw1,dw2}, in which unification of gauge couplings and a large top Yukawa are both realized at the same time. In this context the SM gauge degrees of freedom are localized in a 4-cycle $S_{\rm GUT}$ of the internal dimensions, while chiral matter fields in the $\mathbf{10}$ or $\mathbf{\bar{5}}$ representations are localized at certain 2-cycles of $S_{\rm GUT}$. Finally, Yukawa couplings are generated at the points of intersection of such matter curves, and can be computed via the overlap integral of the internal wavefunctions for these chiral fields. In fact, because this integral is dominated by the wavefunction profiles around the Yukawa point $p$, only the information in a small region around $p$ is necessary to understand the general features of Yukawa couplings in local F-theory models. This in principle allows to perform detailed computations and to obtain universal results for the flavor structure of Yukawa couplings in F-theory GUT's, irrespective of most of the data that describe such models. 

One general lesson that has already been drawn is that fermion mass hierarchies can be easily obtained by restricting the number of Yukawa points. Indeed, as shown in \cite{cchv09} (see also \cite{hv08,fi1,cp09}) the matrix of down-like Yukawa couplings will have rank one if there is a single $\mathbf{10 \times \bar{5} \times \bar{5}}$ Yukawa point $p_{\rm down }$, with a similar statement for up-like Yukawas. This automatically gives a flavor structure in which one family of fermions is much heavier than the other two, whose masses can be generated by D3-brane instantons or a gaugino condensate localized in a different 4-cycle of the compactification, along the lines of \cite{mm09}.\footnote{For different approaches to the generation of fermion mass hierarchies in F-theory see e.g.  \cite{fi08,DudasPalti,Ross,Krippendorf}.}

\begin{figure}[ht]
\begin{center}
\includegraphics[width=8cm]{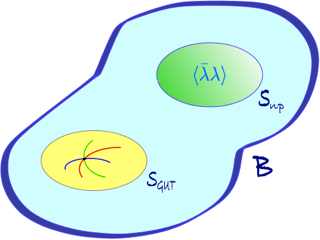}
\end{center}
\caption{\small{Sources of corrections to 7-brane Yukawas in the scenario of \cite{mm09}. The Yukawas on a 7-brane stack wrapping the four-cycle $S_{GUT}$ are modified by the gaugino condensate on 7-branes on a different four-cycle $S_{\rm np}$.}}
\label{backr}
\end{figure}

A detailed analysis of this scenario was performed in \cite{fimr12} (see also \cite{afim}) for the $\mathbf{10 \times \bar{5} \times \bar{5}}$ down-like Yukawas of local F-theory SU(5) models. It was found that non-perturbative effects distort the wavefunction profile of the wavefunction near the Yukawa point in a rather non-trivial way, and that this generates a hierarchy of fermion mass eigenvalues of the form $(1, \eps, \eps^2)$, with $\eps$ a small parameter that measures the size of the non-perturbative effects. Such hierarchy is already present at the level of holomorphic Yukawa couplings which depend on very few parameters of the F-theory model. The physical Yukawa couplings, on the other hand, depend on more detailed information of the local F-theory model, and in particular on the hypercharge flux $F_Y$ which is the agent necessary to break the SU(5) gauge symmetry down to the $SU(3) \times SU(2) \times U(1)_Y$. This latter dependence allows to explain why at the unification scale the Yukawas for the leptons are larger than those of D-quarks. 

The purpose of this work is to extend this previous analysis and apply the scenario proposed in \cite{mm09} to the computation of up-like $\mathbf{10} \times \mathbf{10} \times \mathbf{5}$ Yukawa couplings. The computation of such couplings is already involved in the absence of non-perturbative effects, because of the non-trivial local geometry that is associated to such couplings. Indeed, it has been shown that in order to reproduce the desired rank one structure one must either take into account the phenomenon of 7-brane monodromy \cite{hktw} or to describe this Yukawa point via non-Abelian 7-brane profiles \cite{hktw2,cchv10}, dubbed T-branes in the second reference. In this paper we will take the latter approach and compute up-like Yukawa couplings for a T-brane background in the presence of non-perturbative effects, merging the setups of \cite{mm09} and \cite{cchv10}. As we will see, one again obtains the hierarchical structure $(1, \eps, \eps^2)$ for up-like fermion masses when both setups are combined. Again, this hierarchy arises at the level of holomorphic Yukawa couplings and permeates to the eigenvalues of the physical Yukawa matrix, which we compute via wavefunction overlap. We then show that for a reasonable choice of local model parameters one may obtain a $\CO(1)$ Yukawa for the top quark, justifying the initial motivation that led to consider F-theory SU(5) models as opposed to their type IIB cousins. Finally, we also show how the above hierarchical structure allows to accommodate the observed ratios of U-quark masses, by using quite similar parameters to those necessary to accommodate D-quark and lepton masses in the same SU(5) scheme.

This paper is organized as follows. In Section \ref{s:review} we review the construction of local F-theory GUT's, with emphasis on the geometry that describes up-like Yukawa couplings. In Section \ref{s:E6} we construct a local $E_6$ model in which such Yukawas are generated at a single point. In Section \ref{s:holoyuk} we perform a residue computation in order to compute the holomorphic Yukawa couplings of such local model, showing that the presence of non-perturbative effects generates a rank 3, hierarchical Yukawa matrix. In Section \ref{s:zeromodes} we compute the explicit wavefunction profile for the chiral zero modes of this model, and in particular the corrections to such wavefunctions due to the presence of non-perturbative effects. In section \ref{s:physyuk} we use such wavefunctions to compute the matrix of physical up-like Yukawas, matching the results obtained via residues, and discuss how such Yukawa structure allows to reproduce U-quark masses that are consistent with current experimental data. Final comments and conclusions are left for Section \ref{s:conclu}.

Several technical details have been relegated to the appendices. Appendix \ref{ap:wave} contains details in the computation of zero mode wavefunctions and their corrections due to non-perturbative effects. Appendix \ref{ap:chiral} discusses the choices of fluxes on the local $E_6$ model motivated by the concept of local chirality and doublet-triplet splitting. Appendix \ref{ap:ellip} describes the geometry of the elliptic fibration that is associated to the local  $E_6$ model.

\section{Non-perturbative effects in local F-theory models}
\label{s:review}

One of the most interesting features of GUT models in F-theory is that they naturally lead to a bottom-up approach \cite{aiqu} for building realistic string theory vacua. In particular, in order to analyze the GUT gauge sector of the 4d effective action one just needs to describe the F-theory model in a local patch of the compactification manifold, namely around a 4-cycle $S_{\rm GUT}$ where all the fields charged under the GUT gauge group are localized. In the following we review the basic features of such local F-theory models, with particular emphasis to the geometry that describes the generation of up-like Yukawa couplings. We also review why, in models with hierarchical fermion masses, the presence of non-perturbative effects is necessary in order to obtain a realistic pattern of Yukawas, and how the inclusion of those non-perturbative effects can be made compatible with the above local approach in the same spirit as \cite{mm09,afim,fimr12}.

\subsection{Local F-theory models and up-type Yukawas}

In the standard approach to F-theory GUT model building \cite{bhv1,bhv2,dw1,dw2} (see \cite{ftheoryreviews} for reviews) one considers an elliptic fibration on a threefold base $B$ such that the fiber singularity type over a 4-cycle $S_{\rm GUT}$ corresponds to the desired GUT gauge group $G_{\rm GUT}$. For the purpose of analyzing the gauge theory related to $G_{\rm GUT}$ one may then focus on the region of $B$ that contains $S_{\rm GUT}$, which in the bottom-up terminology of \cite{aiqu} is described as building a local F-theory model. 

A crucial feature of these local geometries is that the fiber singularity on the {\em bulk} of $S_{\rm GUT}$ must correspond to a Dynkin diagram such that the related Lie group is $G_{\rm GUT}$. However, on certain complex submanifolds of $S_{\rm GUT}$ one may have that the fiber singularity is enhanced and corresponds to a higher rank gauge group containing $G_{\rm GUT}$. In particular, on certain curves $\Sigma_i \subset S_{\rm GUT}$ the singularity will correspond to the groups $G_{\Sigma_i} \supset G_{\rm GUT}$. This geometry is usually interpreted in terms of a stack of $(p,q)$ 7-branes wrapping $S_{\rm GUT}$ and generating the gauge group $G_{\rm GUT}$, as well as additional 7-branes wrapping divisors $S_i \subset B$ such that $\Sigma_i = S_i \cap S_{\rm GUT}$. Just like in type IIB, the intersection curves $\Sigma_i$ (dubbed matter curves of $S_{\rm GUT}$) will localize matter fields charged under $G_{\rm GUT}$. The representation of the matter field can be read from the enhanced group $G_{\Sigma_i}$: if we consider $G_{\rm GUT} = SU(5)$, then matter curves $\Sigma_{\mathbf{10}}$ with enhancement to $SO(10)$ will contain matter in the $\mathbf{10}$ or $\mathbf{\overline{10}}$ representations, while curves $\Sigma_{\mathbf{5}}$ with enhancement to $SU(6)$ will contain matter in the $\mathbf{5}$ or $\mathbf{\bar{5}}$. Finally, when two or more of these matter curves meet at a point $p$ there will be further enhancement to a group $G_p$, which signals the appearance of a Yukawa interaction between the matter fields of the curves meeting at $p$. Again, the enhanced group $G_p$ tells us which kind of Yukawa coupling is being developed at this point. For $G_{\rm GUT} = SU(5)$, down-like $\mathbf{10 \times \bar{5} \times \bar{5}}$ Yukawa couplings correspond to points of $SO(12)$ enhancement, while for up-like $\mathbf{10 \times 10 \times 5}$ Yukawas we expect an enhancement to $E_6$.

As pointed out in \cite{bhv1,bhv2,dw1,dw2} an alternative description of these local models can be given in terms of a 8d action related to the 7-branes wrapping $S_{\rm GUT}$ and those intersecting them. This 8d action is defined on a 4-cycle $S$, on which we need to perform dimensional reduction in order to obtain the effective 4d gauge theory. In this sense the Yukawa couplings between 4d matter fields can be computed from the superpotential
\be
W\, =\, m_*^4 \int_{S} \tr \left( F \wedge \Phi \right)
\label{supo7}
\ee
where $m_*$ is the F-theory characteristic scale, $F = dA - i A \wedge A$ is the field strength of the 7-branes gauge boson $A$, and $\Phi$ is the so-called Higgs field: a (2,0)-form on the 4-cycle $S$ describing the 7-branes transverse geometrical deformations. Both $A$ and $\Phi$ transform in the adjoint of a non-Abelian gauge group $G$ that contains $G_{\rm GUT}$ and $G_{\Sigma_i}$, which for the purposes of analyzing Yukawa couplings at $p$ it can be taken to be $G_p$. Finally, the dynamics of this system is also encoded in the D-term
\be
D\, =\, \int_S \omega \wedge F + \frac{1}{2}  [\Phi, \bar{\Phi}]
\label{FI7}
\ee
where $\omega$ stands for the fundamental form of $S$. These two functionals determine the conditions that the 7-branes must satisfy in order to have a stable local F-theory model, as well as the equations of motion for the 7-brane zero mode fluctuations.

From this perspective the presence of matter curves and Yukawa points is understood in terms of the background profile $\langle \Phi \rangle$ for the Higgs field, which in the absence of worldvolume fluxes $F$ depends holomorphically on the complex coordinates of $S$. At a generic point of $S$ this background will only commute with the generators of the subgroup $G_{\rm GUT} \times \prod_a U(1)_a \subset G_p$. At particular complex curves $\Sigma_i$ the rank of $\langle \Phi \rangle$ will jump down and there will be an enhancement of the commutant group, signaling the presence of matter fields localized at such curves. Finally, the commutant group will be maximal at the point where the matter curves meet, namely the Yukawa point $p$. 

In Section \ref{s:E6} we will describe a local F-theory model with $G_p = E_6$ and $G_{\rm GUT} = SU(5)$ precisely from this perspective. As will be illustrated there another important ingredient of the model is a background profile $\langle A \rangle$ for the 7-brane gauge boson or in other words a 7-brane worldvolume flux $\langle F \rangle$ along $S$. The presence of this worldvolume flux is important for two reasons. First it creates a 4d chiral spectrum, selecting a 4d chirality for the zero modes at a given matter curve. An important feature of these 4d chiral modes is that their internal wavefunction profile is non-trivial along the matter curves $\Sigma_i$, and so typically they are fully localized on a particular neighborhood of the GUT 4-cycle. Second, it allows to break the gauge group as $G_{\rm GUT} \raw G_{\rm MSSM}$ by switching on a component of the flux along the hypercharge generator $Q_Y$ \cite{bhv2}. 

In a nutshell, describing the F-theory local model in terms of the 7-brane 8d action allows to encode the local model data in terms of the background profiles $\langle \Phi \rangle$ and $\langle A \rangle$. The 4d chiral modes are described in terms of their internal wavefunction profiles, and the Yukawa couplings between these modes in terms of their overlapping integrals. Using these ingredients one may argue that Yukawas can be computed by simply looking at a region of $S$ near a Yukawa point $p$. Moreover, it was proposed in \cite{hv08} an scenario where all up-like Yukawas are generated from a single Yukawa point $p_{\rm up}$, and all down-like Yukawas from $p_{\rm down}$, in order to obtain a hierarchical pattern of fermion masses. In order to compute up-like (or down-like) Yukawa couplings in such scenario one may then only describe the F-theory GUT model in the vicinity of a single point. This ultra-local approach has been pursued in \cite{hv08,fi09,cchv09,cchv10,afim,palti12,fimr12} and it is also the one followed here in order to compute up-like Yukawa couplings. 

Describing up-like Yukawa couplings ultra-locally involves an important subtlety with respect to describing down-like Yukawas. Namely, in a $\mathbf{10 \times 10 \times 5}$ point it may be the case that two different $\mathbf{10}$ matter curves meet. If this is so the pattern of fermion masses will display the wrong hierarchy, with two heavy and one light families. It was however pointed out in \cite{hktw} that two $\mathbf{10}$ curves that locally seem different may be understood as two branches of the same smooth curve $\Sigma_{\mathbf{10}}$ by taking into account the phenomenon of 7-brane monodromy and that, precisely when this happens, up-like Yukawas are developed at the intersection of $\Sigma_{\mathbf{10}}$ and a $\mathbf{5}$ matter curve $\Sigma_{\mathbf{5}}$, with just one heavy family of up-type quarks, in agreement with empirical data. Unfortunately, a local wavefunction analysis for this sort of geometry has proven to be challenging \cite{hktw,hktw2}.

In this work we would like to analyze up-like Yukawa couplings from a different approach, namely following the proposal in \cite{cchv10} to realize $\mathbf{10 \times 10 \times 5}$ couplings by means of T-brane configurations.\footnote{See \cite{hktw2} for a previous analysis of F-theory models with a non-Abelian Higgs background.} As discussed in \cite{cchv10}, F-theory models based on T-brane backgrounds generalize the concept of 7-brane monodromy and allow to develop up-like Yukawas that result in just one heavy family of up-type quarks. A characteristic feature of T-branes is the fact that the Higgs profile $\langle \Phi \rangle$ does not necessarily commute with other elements of the background, and in particular we have that $[\langle \Phi \rangle, \langle \overline{\Phi}\rangle] \neq 0$. As a result, in order to satisfy the D-term (\ref{FI7}) a compensating non-primitive background flux $\langle F_{\rm np} \rangle$ needs to be switched on, unlike in standard models of intersecting 7-branes. Such background fluxes will satisfy complicated differential equations which in simple T-brane examples like the ones considered in \cite{cchv10} and in this paper can be related to the Painlev\'e equation of the third kind (see \cite{cftw11} for more involved systems). This will of course complicate the analysis, but as we will see within the ultra-local approach one can still solve for the zero mode wavefunctions and compute the matrix of physical Yukawa couplings. 

As a first application of our results one can verify that the resulting pattern of up-like Yukawa couplings indeed reproduces just one heavy family of up-like quarks. In fact, as could be advanced from the results of \cite{cchv10} the up-like Yukawa mass matrix is exactly of rank one, and so two families of quarks are massless. To circumvent this rank one problem one must implement the proposal in \cite{mm09} and take into account how external non-perturbative effects contribute to the Yukawa couplings, as we discuss in the following.


\subsection{Adding non-perturbative effects}

In addition to the divisor $S_{\rm GUT}$ and $S_i$ that describe the local GUT model, a global F-theory compactification will contain other set of divisors of the threefold base $B$ that are also wrapped by branes. Typical examples are hidden sector 7-branes that develop a gaugino condensate, or Euclidean 3-branes that contribute to the superpotential of the 4d effective theory. As pointed out in \cite{mm09} the non-perturbative effects sourced by these sector will also contribute non-trivially to the Yukawa couplings of a local F-theory GUT model, by adding a contribution to the superpotential (\ref{supo7}) that allows to increase the rank of the Yukawa matrix from one to three. 

The basic idea of \cite{mm09} is that non-perturbative effects in a 4-cycle $S_{\rm np} \subset B$ will generate a superpotential of the form
\be
W_{\rm np}\, =\, m_*^3\, e^{-f_{\rm np}}\, =\, m_*^3\, e^{-T_{\rm np} - f_{\rm np}^{\rm 1-loop}}
\ee
where $T_{\rm np} = \int_{S_{\rm np}} J^2 + i C_4$ is the gauge kinetic function of a 7-brane wrapping $S_{\rm np}$ computed at tree-level, and $f_{\rm np}^{\rm 1-loop}$ contains threshold corrections. These corrections will depend on 4d gauge invariant operators that involve 7-brane fields, and in particular they could depend on the Yukawa couplings of the GUT sector, as the results of \cite{ag06} already hint. 

One can in fact check this claim explicitly for F-theory local models because, following the computations in \cite{mm09}, one arrives to the expression
\be
f^{\rm 1-loop}_{{\rm np}}\, =\, - \, {\rm log \, } \ca \, - \frac{1}{8\pi^2} \int_S \str ({\rm log\, } h\, F \wedge F)
\label{f1loopa}
\ee
where $\ca$ depends on the bulk moduli of the three-fold base $B$ and all the dependence of the GUT 7-brane fields is encoded in the integrand. The fact that the superpotential is sourced from $S_{\rm np}$ is encoded in the presence of $h$, which is the holomorphic divisor function of this 4-cycle $S_{\rm np} = \{h = 0\}$. Expanding this expression as in \cite{mm09}  one finally obtains
\be
W_{\rm np} =\, m_*^4\left[  \frac{\eps}{2} \sum_{n \in \IN} \int_S  \theta_n \,  \str \left( \Phi_{xy}^n F \wedge F\right)\right]
\label{suponpintro}
\ee
where 
\be
\eps \, =\, \CA\, e^{-T_{\rm np}} h_0^{N_{\rm D3}}
\label{epsilon}
\ee
with $h_0 = \int_S h$ and $N_{\rm D3} = (8\pi^2)^{-1} \int_S \tr (F \wedge F)\in \IN$, and where $\theta_n$ is proportional to the $n^{\rm th}$ derivative of $h$ normal to $S$. See Appendix C of \cite{fimr12} for explicit expressions of these quantities and a detailed derivation of (\ref{suponpintro}). 

Notice that (\ref{suponpintro}) is written in terms of the GUT 7-brane fields $\Phi$ and $A$, just like the tree-level superpotential (\ref{supo7}). As a result one can add up both expressions and apply the ultra-local approach to compute 7-brane zero mode wavefunctions near a Yukawa point. This computation was carried out in \cite{fimr12} for the down-type Yukawa point $p_{\rm down}$ obtaining that thanks to (\ref{suponpintro}) a hierarchical, rank 3 matrix of Yukawas is generated. In fact, it was found that this effect is already captured by the first term of the sum in (\ref{suponpintro}), namely the term that depends on $\theta_0$ and which is the least suppressed in the derivative expansion of $h$. Hence, in order to see if non-perturbative effects solve the rank 1 problem for up-type Yukawa couplings one may simply consider this first term in the derivative expansion of $W_{\rm np}$ and write the corrected superpotential
\be
W_{\rm total} =\, m_*^4\,  \int_S \tr \left( F \wedge \Phi \right) + \eps\, \frac{\theta_0}{2}   \tr \left( F \wedge F\right)
\label{suponp0intro}
\ee
where $\theta_0\, =\,  (4\pi^2 m_*)^{-1} [{\rm log\, } h/h_0]_{z=0}$. Finally, as shown in Appendix C of \cite{fimr12} these non-perturbative effects do not correct the 7-brane D-term. Hence (\ref{FI7}) and (\ref{suponp0intro}) will be the two expressions in which our local wavefunction analysis will be based.

\section{The $E_6$ model}
\label{s:E6}

In the following we describe the $E_6$ local F-theory model which will serve to compute up-type quark Yukawa couplings. Similarly to the $SO(12)$ model of \cite{fimr12}, one may first consider the 7-brane Higgs background that defines the structure of matter curves and breaks the $E_6$ symmetry down to $SU(5)$, and then describe the background 7-brane flux  that induces 4d chirality and breaks the GUT spectrum down to the MSSM. 

Unlike in the $SO(12)$ case the Higgs background will be in part specified by a T-brane configuration and, as mentioned above, this implies that the Higgs and flux backgrounds are related by the equations of motion. As we will see in section \ref{s:zeromodes} this feature of T-branes will have a direct impact on the zero mode wavefunctions localized at the matter curves, and this will in turn affect the physical Yukawa couplings computed in section \ref{s:physyuk}.

\subsection{Matter curves near the $E_6$ point}

In the standard framework of $SU(5)$ local F-theory models, $\mathbf{10} \times \mathbf{10} \times \mathbf{5}$ Yukawa couplings are developed at points $p$ where an enhanced $E_6$ symmetry occurs. This implies that in order to compute such Yukawas we must consider a 7-brane action where the fields $\Phi$ and $A$ take values in the adjoint of $E_6$. Both $\Phi$ and $A$ will have non-trivial background profiles along the 4-cycle $S$, and so the gauge symmetry group will not be $E_6$ but a subgroup that commutes with both $\langle \Phi \rangle$ and $\langle A \rangle$ at any point of $S$. 

A local $SU(5)$ model with $E_6$ enhancement, dubbed $E_6$ model in the following, can be described by specifying the profiles $\langle \Phi \rangle$ and $\langle A \rangle$ in the vicinity of a $\mathbf{10} \times \mathbf{10} \times \mathbf{5}$ Yukawa point.  By construction, $\langle \Phi \rangle$ and $\langle A \rangle$ are functions of $S$ valued in the Lie algebra of $E_6$, and $\langle \Phi \rangle$ is such that at a generic point of this neighborhood it breaks the $E_6$ symmetry down to $SU(5) \times U(1)^n$, with $n=0,1,2$. Then, by neglecting the effect of the worldvolume flux $\langle A \rangle$, we can identify $G_S = SU(5)$ as the GUT gauge group of this model. 
In addition, the profile $\langle \Phi \rangle$ will describe
the different matter curves, that is the curves of $S$ at which chiral modes in the representations $\mathbf{5}$ or $\mathbf{10}$ are localized.

This picture can be understood in more detail by expressing the local model data in terms of the generators $Q_\a$ of $E_6$. These generators can be decomposed as $\{ Q_\a \}=\{ H_i, E_\rho\}$, where $H_i$ generate the Cartan subalgebra of $E_6$ and $E_\rho$ correspond to the roots of $E_6$. More precisely we have the usual relation
\be
[H_i, E_\rho]\, =\, \rho_i E_\rho
\ee
where $\rho_i$ is the $i$-th component of the root $\rho$. The 72 non-trivial roots are given by
\be
(0,\underline{\pm1,\pm1,0,0,0})
\ee
where we should consider all possible permutation of the underlined vector entries, and
\be
 \oh (\pm \sqrt{3},\pm 1,\pm 1,\pm 1,\pm 1,\pm 1) \quad \quad \mathrm{ with\ even\  number\ of\,} +'{\rm s}
\ee

Near the up-type Yukawa point one can decompose the background profile of $\Phi$ as a linear combination of the above generators, with arbitrary functions of the 4-cycle $S$ as coefficients. If we parametrize the complex coordinates of $S$ as $(x,y)$ then we have that $\Phi= \Phi_{xy}\, dx \wedge dy$ and so in general
\be
\langle \Phi_{xy} \rangle \, =\, \sum_i g_\a \, Q_\a 
\label{genPhi}
\ee
with $g_\a \equiv g_\a(x,\bar{x},y,\bar{y})$ functions in the vicinity of the Yukawa point and $Q_i \in \{ H_i, E_\rho\}$. For simplicity, the generators $Q_\a$ are often chosen to lie within the Cartan subalgebra of $E_6$, because then one can understand the background (\ref{genPhi}) as a configuration of intersecting 7-branes. For instance, one may consider the following background
\be
\langle \Phi_{xy} \rangle \, =\, m^{3/2} \sqrt{x}\, P + \mu^2 \left( bx - y \right) Q
\label{monoPhi}
\ee
where $m$ and $\mu$ are real parameters with the dimension of mass, $b$ is a complex adimensional parameter and $P$ and $Q$ are the following combinations of Cartan generators
\bea
\label{genP}
P &= &\frac{1}{2}(\sqrt 3 H_1+H_2+H_3+H_4+H_5+H_6)\\
Q & = &\frac{1}{2}\Big( \frac{5}{\sqrt{3}} H_1 -H_2-H_3-H_4-H_5-H_6\Big)
\label{genQ}
\eea

Given a background (\ref{genPhi}) one can analyze the symmetry breaking pattern of the local model and understand the structure of its matter curves \cite{bhv1,bhv2}. The basic quantity to look at is $[\langle \Phi_{xy} \rangle, E_\rho]$, which will be a function valued on the Lie algebra of $E_6$ and tells us to which subgroup the initial $E_6$ group is broken. For instance, for the background (\ref{monoPhi}) the set of generators that commute with $\langle \Phi_{xy} \rangle$ for all points of $S$ is the set of roots
\be
(0,\underline{1,-1,0,0,0})
\label{rootsSU5}
\ee
as well as the Cartan generators. This implies that the subgroup of $E_6$ that remains as a gauge symmetry group is given by $SU(5) \times U(1)^2$, and the GUT gauge group can be identified with $G_S = SU(5)$. 

At particular submanifolds of $S$ there will be extra sets of roots that commute with $\langle \Phi_{xy} \rangle$, implying an enhancement of the bulk symmetry group. In particular we have that there is such enhancement for two different holomorphic curves, namely
\bea
\label{mono5}
\Sigma_{\mathbf{5}} = \{ bx-y=0\} & \raw & \pm\oh ( \sqrt{3},\underline{1,-1,-1,-1,-1})\\ \label{mono10}
\Sigma_{\mathbf{10}} = \{\mu^4 (bx-y)^2 = m^3 x\} & \raw & \pm (0,\underline{1,1,0,0,0}) \\  \nonumber
& {\rm or} & \, \pm \oh (-\sqrt{3},\underline{1,1,-1,-1,-1})
\eea
where at the lhs we have displayed the matter curve or curve of enhancement and at the rhs the extra roots that commute with $\langle \Phi_{xy} \rangle$ at such curve. At the curve (\ref{mono5}) there are ten additional roots that together with (\ref{rootsSU5}) and the Cartan subalgebra generate the group $SU(6) \times U(1)$. These extra roots transform as either a $\mathbf{5}$ or a $\mathbf{\bar{5}}$ representation of $SU(5)$, and so will the zero modes that are localized there \cite{bhv1,bhv2}. Following the common practice one then dubs $bx-y=0$ as the $\mathbf{5}$ matter curve $\Sigma_{\mathbf{5}}$ of the local model. At the curve (\ref{mono10}) there are 20 extra unbroken roots transforming in the representations $\mathbf{10}$ and $\mathbf{\overline{10}}$ of $SU(5)$, enhancing the bulk symmetry group to $SO(10) \times U(1)$ and giving rise to a $\mathbf{10}$ matter curve $\Sigma_{\mathbf{10}}$. Finally, at the intersection point $p_{\rm up} = \{x=y=0\}$ of both curves $\langle\Phi_{xy}\rangle = 0$, and so the full $E_6$ symmetry remains unbroken. It is at this point where a Yukawa $\mathbf{10} \times \mathbf{10} \times \mathbf{5}$ must be generated via triple overlap of zero mode wavefunctions. 

The $\mathbf{10}$ curve (\ref{mono10}) requires some further explanation, as the roots that enhance the symmetry are not the same all over it. Indeed, at the branch $\sqrt{x} = bx-y$ we have that the roots in the first line of (\ref{mono10}) are the ones that commute with the background, while for $-\sqrt{x} = bx-y$ the roots of the second line are the ones commuting with $\langle \Phi_{xy} \rangle$. While this make look puzzling, it was realized in \cite{hktw} that the zero modes of the two branches of the $\mathbf{10}$ curve (\ref{mono10}) are identified by the phenomenon of 7-brane monodromy. In fact, it was also pointed out in \cite{hktw} that such monodromy is necessary in order to achieve precisely one heavy generation of up-type quarks whenever $\langle \Phi_{xy} \rangle$ takes values in the Cartan of $E_6$, and a background similar to (\ref{monoPhi}) was proposed as a candidate to obtain realistic up-like Yukawas. However, the analysis in \cite{hktw,hktw2} shows that it is not obvious to find non-singular solutions for the zero mode wavefunctions near the intersection point of matter curves in such monodromic 7-brane configurations. As this is the region of larger wavefunction overlap and the one that contributes most to the value of the Yukawa couplings, this complicates the computational and predictive power of such local model. 

One can however consider an alternative background for the transverse position field $\Phi$, based on the proposal made in \cite{cchv10} of describing up-like Yukawa couplings via T-branes. Indeed, let us consider the background
\be
\langle \Phi_{xy} \rangle =m( E^+ + m\,x E^-) +\mu^2 (bx - y) Q
\label{TbranePhi}
\ee
where all quantities are as in (\ref{monoPhi}) except for the generators $E^{\pm}$ whose corresponding roots, also denoted $E^\pm$, are defined as
\be
E^\pm =\pm \oh (\sqrt{3},1,1,1,1,1)
\label{Epm}
\ee
and satisfy the relation $[E^+,E^-]=P$. More precisely, the triplet $\{E^+,E^-,P\}$ generates the $\mathfrak{su}(2)$ factor of a $\mathfrak{su}(5)\oplus\mathfrak{su}(2)\oplus\mathfrak{u}(1)$ maximal Lie subalgebra of $\mathfrak{e}_6$, under which the $E_6$ adjoint decomposes as
\begin{equation}
\mathbf{78} \rightarrow (\mathbf{24,1})_0 \oplus (\mathbf{1,3})_0 \oplus (\mathbf{1,1})_0 \oplus (\mathbf{10,2})_{-1} \oplus (\mathbf{\overline{10},2})_{1}  \oplus (\mathbf{5,1})_2 \oplus (\mathbf{\overline{5},1})_{-2}
\label{decomp78}
\end{equation}
From this decomposition it is manifest that the pair of $\mathbf{10}$'s described above transform as a doublet of the $SU(2)$  generated by $\{E^+,E^-,P\}$. In particular if we define
\be
\begin{array}{rcl}
E_{\mathbf{10}^+} = (0,\underline{1,1,0,0,0}) & \quad & E_{\mathbf{10}^-} = \oh (-\sqrt{3},\underline{1,1,-1,-1,-1}) \\
E_{\mathbf{\overline{10}}^+} = - (0,\underline{1,1,0,0,0}) & \quad & E_{\mathbf{\overline{10}}^-} = - \oh (-\sqrt{3},\underline{1,1,-1,-1,-1}) \\
\end{array}
\label{10roots}
\ee
we have the relations
\be
[E^\pm ,E_{\mathbf{10}^\mp}] = E_{\mathbf{10}^\pm}, \quad \quad [E^\pm ,E_{\mathbf{10}^\pm}] = 0, \quad \quad[ P, E_{\mathbf{10}^\pm} ]= \pm E_{\mathbf{10}^\pm}.
\label{Edoublet}
\ee

Let us analyze the gauge symmetry group of this background and the structure of matter curves. Just as in the previous case we have to look at the commutant of $\langle \Phi\rangle$ as a function of the coordinates $x,y$. The gauge group is the commutant at generic points while the matter curves are identified by finding jumps in its rank \cite{cchv10}.  For the background (\ref{TbranePhi}) one can easily check that the set of roots of the subalgebra $\mathfrak{su}(5)\oplus\mathfrak{u}(1) \subset \mathfrak{e}_6$ do commute at generic points in $S$ and so we can identify the GUT gauge group with $SU(5)$.

Regarding the matter curves, we find that at $\Sigma_{\mathbf{5}}=\{bx-y=0\}$ the roots $(\mathbf{5,1})_2 = \oh ( \sqrt{3},\underline{1,-1,-1,-1,-1}) = E_{\mathbf{5}}$ and $(\bar {\mathbf{5}},\mathbf{1})_{-2} = \oh (-\sqrt{3},\underline{-1,1,1,1,1}) = E_{\bar{\mathbf{5}}}$ commute with $\langle \Phi\rangle$, since
\bea
\, [\langle \Phi\rangle, E_{\mathbf{5}}] & = & 2\mu^2 (bx-y) E_{\mathbf{5}} \\
\, [\langle \Phi\rangle, E_{\bar{\mathbf{5}}}] & = & -2\mu^2 (bx-y) E_{\bar{\mathbf{5}}}
\eea
and so at $\Sigma_{\mathbf{5}}$ the symmetry group enhances to $SU(6)\times U(1)$.
Similarly, the action of $\langle \Phi\rangle$ on the sector $(\mathbf{10},\mathbf{2})_{-1}$ is given by 
\be\label{com10}
[\langle \Phi\rangle, R_+E_{\mathbf {10}^+}+R_-E_{\mathbf {10}^-}] = 
\left(
\begin{array}{cc}
-\mu^2(bx-y) & m \\ m^2 x & -\mu^2(bx-y)
\end{array}
\right)
\left(
\begin{array}{c}
R_+ E_{\mathbf {10}^+} \\ R_-E_{\mathbf {10}^-}
\end{array}
\right)
\ee
while for the conjugate sector $(\overline{\mathbf{10}},\mathbf{2})_{1}$ we have
\be\label{com10b}
[\langle \Phi\rangle, R'_+ E_{\mathbf{\overline{10}}^+}+R'_-E_{\mathbf{\overline{10}}^-}] = 
\left(
\begin{array}{cc}
\mu^2(bx-y) & -m^2 x \\ -m & \mu^2(bx-y)
\end{array}
\right)
\left(
\begin{array}{c}
R'_+E_{\mathbf{\overline{10}}^+} \\ R'_-E_{\mathbf{\overline{10}}^-}
\end{array}
\right)
\ee
where $R_\pm,\,R'_\pm$ are functions on $S$. At $\Sigma_{\mathbf{10}} = \{\mu^4 (bx-y)^2 = m^3 x\}$ the matrices in (\ref{com10}) and (\ref{com10b}) have vanishing determinant so there are additional roots commuting with $\langle \Phi\rangle$, and therefore a jump in the rank of the symmetry group.\footnote{At $\Sigma_{10}$ both roots $R\equiv E_{\mathbf{10}^+}+\frac{\mu^2}{m}(bx-y)E_{\mathbf{10}^-}$ and $R'\equiv\frac{\mu^2}{m}(bx-y) E_{\mathbf{\overline{10}}^+}+E_{\mathbf{\overline{10}}^-}$ commute with $\langle\Phi\rangle$ but since these are not conjugate to each other the enhanced algebra is a complex subalgebra of $\mathfrak{e}_6^{\mathbb C}$ that is not the complexification of a real algebra. Thus, we cannot associate a real gauge group to the matter curve $\Sigma_{\mathbf{10}}$ in agreement with the discussion in section 4.1 of \cite{cchv10}.} We therefore identify $\Sigma_{10}$ with the $\mathbf{10}$ curve of this T-brane background. Notice that we arrive to the same matter curves $\Sigma_{\mathbf{5}}$, $\Sigma_{\mathbf{10}}$ if we consider the action $[\langle \bar{\Phi}\rangle, \cdot]$, and that as before they both meet at the Yukawa point $p_{\rm up} = \{x=y=0\}$.\footnote{Note that for this local model $\langle \Phi_{xy} \rangle \neq 0$ at  $p_{\rm up}$, and so the symmetry group is no longer $E_6$ at the Yukawa point. As discussed in \cite{cchv10} this is a general feature of T-brane configurations, see also Appendix C. 
By abuse of terminology, we will still refer to this point as the $E_6$ point of the local model. \label{e6p}}

Finally, one can further generalize the above T-brane background by considering the following Ansatz
\be
\langle \Phi_{xy} \rangle =m(e^f E^+ + m\,x e^{-f} E^-) +\mu^2 (bx-y) Q
\label{TbranefPhi}
\ee
with $f \equiv f(x,\bar{x},y,\bar{y})$ an arbitrary real function in $S$. It is easy to check that everything works as before, and that we 
recover the same two matter curves $\Sigma_{\mathbf{5}}$ and $\Sigma_{\mathbf{10}}$. As we will see in the following, the more general Ansatz (\ref{TbranefPhi}) is required by the equations of motion for the background, with a very specific choice of real function $f$. 

\subsection{The T-brane background}
\label{ss:Tbranebkg}

When considering a background profile for the 7-brane position field $\Phi$ of the form (\ref{genPhi}), one should make sure that it satisfies the equations of motion that arise form the 7-brane superpotential (\ref{supo7}) and D-term (\ref{FI7}). These read
\begin{subequations}
\label{Fterm7}
\begin{align}
\bar \partial_{A} \Phi = &\, 0\\
F^{(0,2)} =&\, 0
\end{align}
\end{subequations}
for the F-term equations and 
\be
\omega \wedge F + \frac{1}{2} [ \Phi , \bar{\Phi}]=0
\label{Dterm7}
\ee
for the D-term equation. Evaluating these equations at the level of the background, one sees that setting $\langle A \rangle =0$ and choosing $g_\a$ to be holomorphic functions in (\ref{genPhi}) the F-term equations (\ref{Fterm7}) are trivially satisfied. If in addition $\langle \Phi \rangle$ only involves Cartan generators then $[\langle \Phi \rangle, \langle \bar{\Phi} \rangle] = 0$ and the background D-term equation (\ref{Dterm7}) is also satisfied. This sort of configuration is nothing but the standard strategy to build F-theory GUT models, since the above profile for $\langle \Phi \rangle$ corresponds to a set of 7-branes wrapping different divisors of the threefold base $B$. On top of this background we can add non-trivial worldvolume fluxes $\langle F \rangle$ such that eqs.(\ref{Fterm7}) and (\ref{Dterm7}) are still satisfied, which usually corresponds to switching on a worldvolume flux for each of these 7-branes.

However, our previous discussion led us to 7-brane backgrounds of the form (\ref{TbranePhi}), where $\langle \Phi \rangle$ does not lie along the Cartan generators of $E_6$. Because the functions in (\ref{TbranePhi}) are holomorphic, this background does satisfy the F-term eqs.(\ref{Fterm7}) for $\langle A \rangle =0$, but because now $[\langle \Phi \rangle, \langle \bar{\Phi} \rangle] \neq 0$ the D-term is no longer satisfied in this case. Hence, for configurations where $\langle \Phi \rangle$ is not along the Cartan a non-trivial worldvolume flux $\langle A \rangle$ should always be switched on in order for the equations of motion to be satisfied. Notice that this modifies the F-term equations (\ref{Fterm7}), and in fact this prevents to find a simple solution for a 7-brane background with the profile (\ref{TbranePhi}).

Nevertheless, following \cite{cchv10} one can show that the more general Ansatz (\ref{TbranefPhi}) does correspond to a solution to the equations of motion if the appropriate background flux $\langle A \rangle$ is added to it. The basic idea is to realize that the backgrounds (\ref{TbranePhi}) and (\ref{TbranefPhi}) are related by a complexified gauge transformation. These transformation act on the 7-brane fields as
\be
\Phi\ \raw g\,\Phi \, g^{-1}, \qquad A_{0,1} \ \raw \ A_{0,1} + i g\, \bar \partial g^{-1}
\label{cgtbkg}
\ee
where $g$ is obtained by exponentiation of an element of the complexified $\mathfrak{e}_6$ Lie algebra. In particular we can take
\be\label{eq:ans}
g=e^{\frac{f}{2}P}
\ee
with $f$  a real function, so that $f/2 P$ is an element of the complexification of the $\mathfrak{su}(2)$ factor within $\mathfrak{su}(5)\oplus\mathfrak{su}(2)\oplus\mathfrak{u}(1) \subset \mathfrak{e}_6$. Indeed, it is easy to see that acting with (\ref{eq:ans}) on the background (\ref{TbranePhi}) and using that $[P,E^{\pm}]=\pm 2 E^{\pm}$ one obtains (\ref{TbranefPhi}). 

Complexified gauge transformations leave the F-term equations (\ref{Fterm7}) invariant, while the D-term equation (\ref{Dterm7}) transforms non-trivially under them \cite{cchv09}. Hence, starting with a solution to the F-term equations one can produce a new one by acting with (\ref{eq:ans}). A very simple solution of the F-term equations consists in taking $\langle \Phi \rangle$ as in (\ref{TbranePhi}) and $\langle A_{0,1} \rangle = 0$. Acting with (\ref{eq:ans}) on such background one obtains
\be
\langle \Phi_{xy}\rangle \, =\, m(e^{f}E^++mxe^{-f}E^-)+\mu^2 (bx-y) Q, \qquad \langle A_{0,1}\rangle =-\frac{i}{2} \bar{\partial} f P
\label{Tbranebkg}
\ee
which will automatically solve F-term equations, while the D-term equations will constrain the function $f$. Reversing the logic, one could start with a 7-brane background such that $\langle \Phi \rangle$ and $\langle A \rangle$ are specified by (\ref{Tbranebkg}) and $\langle \bar{\Phi}_{xy} \rangle = \langle \Phi_{xy}\rangle ^\dag$, $\langle A_{1,0} \rangle = \langle A_{0,1} \rangle^\dag$. Then, by acting with the inverse of the complexified gauge transformation (\ref{eq:ans}) one can obtain a (non-physical) background in which $\langle {\Phi}_{xy} \rangle$ is given by (\ref{TbranePhi}) and  $\langle A_{0,1} \rangle = 0$. This transformed background is usually dubbed {\em holomorphic gauge} \cite{fi09}, and although non-physical it is a very useful tool to analyze F-term dependent quantities like holomorphic Yukawas, as we will see in the next section. 

In the background (\ref{Tbranebkg}) we have that
\be
[\langle \Phi_{xy} \rangle, \langle \bar{\Phi}_{\bar x\bar y}\rangle]\,=\, m^2(e^{2f}-m^2|x|^2e^{-2f})P \qquad \langle F_{1,1}\rangle \,=\, -i\partial \bar\partial f P
\label{relTbrane}
\ee
and so taking the K\"ahler form to be
\be
\omega=\frac{i}{2}(dx\wedge d\bar x+dy\wedge d\bar y)
\label{kahler}
\ee
we obtain that at the level of the background the D-term equation reads
\be\label{eq:comp}
\left (\partial_x\partial_{\bar x}+\partial_y\partial_{\bar y}\right )f=m^2(e^{2f}-m^2|x|^2e^{-2f})\,,
\ee
which is a rather involved non-linear equation. Nevertheless, switching to polar coordinates $x = r e^{i\theta}$ in the $x$-plane and taking the Ansatz $f = f(r)$ it simplifies to
\be\label{eq:ode}
\left (\frac{d^2}{dr^2}+\frac{1}{r}\frac{d}{dr}\right )f=4m^2(e^{2f}-m^2r^2e^{-2f})\,.
\ee
If we now define the function $h(r)$ such that
\be
e^{2f(r)}=mre^{2h(r)}
\ee
then the equation turns into 
\be
\left ( \frac{d^2}{dr^2}+\frac{1}{r}\frac{d}{dr} \right )h=8m^3r\sinh(2h)\,.
\ee
Finally, if we perform the change of variables $s=\frac{8}{3}(mr)^{3/2}$ we have
\be\label{eq:final}
\left ( \frac{d^2}{ds^2}+\frac{1}{s}\frac{d}{ds} \right ) h =\frac{1}{2}\sinh (2h)
\ee
which is nothing but a particular case of the Painlev\'e III differential equation, more precisely the one found in \cite{cchv10} in the context of T-brane configurations. Solutions to this equation have been found in \cite{mtw77} by requiring that they are bounded at $r \raw \infty$. Since in the present context we are only describing a local patch of the 7-brane configuration we may focus on the asymptotic behavior of the Painlev\'e transcendent near the origin
\be
f(r) =\log c+  c^2 m^2 x\bar{x}+m^4(x\bar{x})^2\left( \frac{c^4}{2}
   -\frac{1}{4 c^2}\right) +\dots\,,
\label{Painori}
\ee
where as in \cite{cchv10} we have imposed regularity of the gauge transformation (\ref{eq:ans}) at $r=0$. Note that the solution is parametrized by a real dimensionless constant $c$, a parameter which should be fixed by the details of the global completion of the 7-brane local model. A natural value for $c$ can be obtained by extending the solution for $f(r)$ to all the real axis and requiring absence of poles. One then obtains \cite{mtw77}
\be
c = 3^{1/3} \frac{\Gamma\left[ \frac{2}{3}\right]}{\Gamma\left[ \frac{1}{3}\right]} \sim 0.73
\label{nopolec}
\ee
where $\G$ is the Gamma function. We would then expect that having no poles in a region around the Yukawa point selects an interval for the possible values for c around (\ref{nopolec}). Fixing the value of $c$ and $m$ fixes the T-brane background of the model, and in particular the non-primitive fluxes in (\ref{relTbrane}) that are necessary to satisfy the D-term equation. 

\subsection{Primitive worldvolume fluxes}

On top of the flux in (\ref{relTbrane}), the above model admits additional contributions to the background worldvolume flux $\langle F \rangle$ if they do not spoil the F-term and D-term conditions. The simplest way to introduce them is to consider primitive $(1,1)$ fluxes $\langle F \rangle$ in the Cartan of $E_6$. Considering such fluxes is important to complete the local F-theory model, not just because they will be  generically there, but also because they play an important role for the phenomenology of the model. On the one hand  they will generate 4d chirality for the $SU(5)$ spectrum, and on the other hand they will break the $SU(5)$ gauge group down to $SU(3) \times SU(2) \times U(1)_Y$.

More precisely, let us consider the worldvolume flux
\be
\langle F_Q \rangle = i \left[ - M(dy\wedge d\bar y-dx\wedge d\bar x) + N(dx\wedge d\bar y+dy\wedge d\bar x)\right]  Q
\label{Qflux}
\ee
where the generator $Q$ is given by (\ref{genQ}), and $M$ and $N$ are flux densities near the Yukawa point that we will approximate by constants. It is easy to check that adding such flux will not spoil the equations of motion for any value of $M$, $N$, which will be considered as real parameters of the model in the following. The presence of such worldvolume flux will induce 4d chirality in the matter curves. Indeed, the modes of opposite chirality {\bf 5}, ${\bf \bar{5}}$ and {\bf 10}, ${\bf \overline{10}}$ feel the  background (\ref{Tbranebkg}) in a similar way, and so whenever there is a zero mode solution for one chirality there will be a solution for the opposite chirality as well. This is no longer true for the background flux (\ref{Qflux}),  that will select locally modes of one chirality or the other depending on the sign of $M$ and $N$. 
A more detailed discussion of the local chirality index can be found in appendix \ref{ap:chiral}.

Besides inducing 4d chirality, worldvolume fluxes break the $SU(5)$ gauge group when switched on along the hypercharge generator \cite{bhv2}. In general realistic GUT F-theory models will have such worldvolume flux, which we can represent locally as
\be
\langle F_Y \rangle = i \left[ \tilde{N}_Y (dy\wedge d\bar y-dx\wedge d\bar x) + N_Y(dx\wedge d\bar y+dy\wedge d\bar x)\right]  Q_Y
\label{Yflux}
\ee
where $N_Y$, $\tilde{N}_Y$ are local flux densities and 
\be
Q_Y\, =\, \frac{1}{3} \left(H_2 + H_3 + H_4\right) - \oh \left(H_5 + H_6\right)
\label{Ygen}
\ee
is the hypercharge generator. Following the common practice we will refer to (\ref{Yflux}) as the hypercharge flux of the local model. This flux will enter into the Dirac equation for the zero modes and, just  as in the local $SO(12)$ model of \cite{fimr12}, it will be the only quantity that will distinguish between particles within the same $SU(5)$ multiplet but with different hypercharge, c.f. table \ref{t:sectors} below.

\subsection{Summary}

Let us summarize the details of the $E_6$ model which we will use to compute up-like Yukawa couplings. If we parametrize the four-cycle $S$ by the complex coordinates $x$, $y$, the Higgs background that breaks $E_6 \raw SU(5) \times U(1)$ is given by 
\be
\langle \Phi_{xy} \rangle =m(e^f E^+ + m\,x e^{-f} E^-) +\mu^2 (bx-y) Q
\label{sumPhi}
\ee
where $m$ and $\mu$ are real parameters with the dimensions of mass, $a$, $b$ are adimensional parameters and $E^\pm$ and $Q$ are the $E_6$ roots given respectively by (\ref{Epm}) and (\ref{genQ}). The real function $f \equiv f(x,\bar{x})$ will solve the equation (\ref{eq:comp}) and  can be approximated  locally by (\ref{Painori}). Finally, one may choose different values for the parameter $b$. For the sake of concreteness when computing physical Yukawas we will restrict to the case 
\be
b\, =\, 1
\label{choiceab}
\ee
although our discussion can be easily generalised to other values of $b$. 

The worldvolume flux of this model will be given by 
\be
\langle F \rangle \, =\, \langle F_{\rm p} \rangle + \langle F_{\rm np} \rangle
\label{sumF}
\ee
where $\langle F_{\rm np} \rangle$ is the non-primitive flux that is necessary to compensate the contribution of $[\langle \Phi_{xy} \rangle, \langle \bar{\Phi}_{\bar x\bar y}\rangle]$ to the D-term equation (\ref{Dterm7}), and reads
\be
\langle F_{\rm np} \rangle\, =\, -i\partial \bar\partial f P
\label{sumFnp}
\ee
with the $E_6$ generator $P$ given by (\ref{genP}). In addition we have that 
\be
\langle F_{\rm p}\rangle = iQ_R(dy\wedge d\bar y-dx\wedge d\bar x)+iQ_S(dx\wedge d\bar y+dy\wedge d\bar x)
\label{sumFp}
\ee
is the primitive flux needed to generate chirality and further break the gauge group as $SU(5) \raw SU(3) \times SU(2) \times U(1)_Y$. Here we have defined 
\be
Q_R=-MQ+\tilde N_YQ_Y,\qquad Q_S=NQ+N_YQ_Y
\label{QRS}
\ee
with $Q_Y$ the hypercharge generator (\ref{Ygen}) and $M$, $N$, $N_Y$, $\tilde{N}_Y$ real flux densities. Because of the presence of the hypercharge generator, zero modes within the same $SU(5)$ multiplet but with different hypercharge will feel a different worldvolume flux, and this will translate into a different internal wavefunction profile for each of them. We have summarized in table \ref{t:sectors} the different sectors that arise in the $E_6$ model together with their charges under the MSSM gauge group and the worldvolume flux operators (\ref{QRS}). The latter charges are defined as
\be
[Q_R, E_\rho]\, =\, q_R \, E_\rho, \quad \quad [Q_S, E_\rho]\, =\, q_S \, E_\rho
\label{QRSch}
\ee
and so are given by a linear combination of flux densities. 

\begin{table}[htb] 
\renewcommand{\arraystretch}{1.2}
\setlength{\tabcolsep}{5pt}
\begin{center}
\begin{tabular}{|c| c|c||c|c|}
\hline
Sector& Root &$G_{\rm MSSM}$  & $q_R$ & $q_S$\\
\hline
\hline
$\mathbf{10}_1$ &$(0,\underline{1,1,0},0,0) \oplus \oh(-\sqrt{3},\underline{1,1,-1},-1,-1) $ &  $(\mathbf{\bar 3}, \mathbf{1})_\frac{2}{3}$ &$ M+\frac{2}{3} \tilde N_Y$& $-N+\frac{2}{3} N_Y$\\ 
\hline
$\mathbf{10}_2$ &$(0,\underline{1,0,0},\underline{1,0}) \oplus \oh(-\sqrt{3},\underline{1,-1,-1},\underline{1,-1}) $ &  $(\mathbf{3}, \mathbf{2})_{-\frac{1}{6}}$ &$ M-\frac{1}{6} \tilde N_Y$ &  $-N-\frac{1}{6} N_Y$\\
\hline
$\mathbf{10}_3$ &$(0,0,0,0,1,1) \oplus \oh(-\sqrt{3},-1,-1,-1,1,1) $ &  $(\mathbf{1}, \mathbf{1})_{-1}$ &  $ M- \tilde N_Y$&  $-N-N_Y$\\
\hline
$\mathbf{5}_1$&$\oh (\sqrt{3},-1,-1,-1,\underline{1,-1})$&$(\mathbf{1}, \mathbf{2})_{-\frac{1}{2}}$ &$-2M-\frac{1}{2}\tilde N_Y$ &$2N-\frac{1}{2} N_Y$\\

\hline
$\mathbf{5}_2$&$\oh (\sqrt{3},\underline{1,-1,-1},-1,-1)$&$ (\mathbf{3}, \mathbf{1})_{\frac{1}{3}}$ &$-2M+\frac{1}{3}\tilde N_Y$ &$2N+\frac{1}{3} N_Y$\\\hline
\hline
$\mathbf{\overline{10}}_1$ &$(0,\underline{-1,-1,0},0,0) \oplus \oh(\sqrt{3},\underline{-1,-1,1},1,1) $ &  $(\mathbf{3}, \mathbf{1})_{-\frac{2}{3}}$ &$ -M-\frac{2}{3}\tilde N_Y$& $N-\frac{2}{3} N_Y$\\ 
\hline
$\mathbf{\overline{10}}_2$ &$(0,\underline{-1,0,0},\underline{-1,0}) \oplus \oh(\sqrt{3},\underline{-1,1,1},\underline{-1,1}) $ &  $(\mathbf{\bar 3}, \mathbf{2})_{\frac{1}{6}}$ &$ -M+\frac{1}{6} \tilde N_Y$ &  $N+\frac{1}{6} N_Y$\\
\hline
$\mathbf{\overline{10}}_3$ &$(0,0,0,0,-1,-1) \oplus \oh(\sqrt{3},1,1,1,-1,-1) $ &  $(\mathbf{1}, \mathbf{1})_{1}$ &  $ -M + \tilde N_Y$&  $N+N_Y$\\
\hline
$\mathbf{\overline{5}}_1$&$\oh (-\sqrt{3},1,1,1,\underline{-1,1})$&$(\mathbf{1}, \mathbf{2})_{\frac{1}{2}}$ &$2M+\frac{1}{2}\tilde N_Y$ &$-2N+\frac{1}{2} N_Y$\\
\hline
$\mathbf{\bar 5}_2$&$\oh (-\sqrt{3},\underline{-1,1,1},1,1)$&$ (\mathbf{\bar 3}, \mathbf{1})_{-\frac{1}{3}}$ &$2M-\frac{1}{3}\tilde N_Y$ &$-2N-\frac{1}{3} N_Y$\\
\hline
\hline
$\mathbf{X}^+,\mathbf{Y}^+$ & $(0, \underline{1,0,0},\underline{-1,0})$ & $(\mathbf{3}, \mathbf{2})_{\frac{5}{6}}$ & $\frac{5}{6}\tilde N_Y$ &$\frac{5}{6} N_Y$ \\
\hline
$\mathbf{X}^-,\mathbf{Y}^-$ & $(0, \underline{-1,0,0},\underline{1,0})$ & $(\mathbf{\bar 3}, \mathbf{2})_{-\frac{5}{6}}$ & $-\frac{5}{6}\tilde N_Y$ &$-\frac{5}{6} N_Y$ \\
\hline
\end{tabular}
\end{center}
\caption{Different sectors and charges for the $E_6$ model.}
\label{t:sectors}
\end{table}

As we will see in section \ref{s:zeromodes}, the quantities $q_R$, $q_S$ enter into the expressions for the internal wavefunctions of the MSSM chiral zero modes. In fact, these charges determine which sectors of those in table \ref{t:sectors} have localized zero modes near the Yukawa point. In order to construct a local model with the MSSM chiral spectrum we need to impose that chiral modes only arise from the four first rows of table \ref{t:sectors}. This will impose some constraints on $q_R$ and $q_S$, which will in turn impose constraints in the values of the flux densities  $M$, $N$, $N_Y$, $\tilde{N}_Y$, as we briefly describe below and in more detail in Appendix \ref{ap:chiral}.

One important constraint comes from avoiding the doublet-triplet splitting problem of 4d $SU(5)$ GUT models. Following \cite{bhv2}, one can do so by adjusting the fluxes so that the sector of triplets $\mathbf{5}_2$, $\mathbf{\overline{5}}_2$ does not feel any net flux and it is then a non-chiral sector without any localized 4d modes. As discussed in Appendix \ref{ap:chiral}, this condition amounts to impose that $q_S(\mathbf{5}_2) =q_S(\mathbf{\bar 5}_2) =0$ or in other words that
\be
N_Y + 6N\, =\, 0
\label{cond23}
\ee
On the other hand, we would like to have a localized chiral mode in the sector $\mathbf{5}_1$ but not in $\mathbf{\overline{5}}_1$. This amounts to require that $q_S(\mathbf{5}_1) > 0$ which, using (\ref{cond23}) translates into
\be
N\, >\, 0
\label{cond23b}
\ee
In addition, we should require that there are localized chiral modes in the sector $\mathbf{10}_i$ but not in $\mathbf{\overline{10}}_i$ for $i=1,2,3$. This can be understood in terms of the condition $q_R(\mathbf{10}_i) > 0$ with is achieved by imposing 
\be
M + q_Y \tilde N_Y \, > \, 0 \quad {\rm for\ }\ q_Y = \frac{2}{3}, -\frac{1}{6}, -1 \qquad \Rightarrow \qquad -\frac{3}{2}<\frac{\tilde N_Y}{M}<6
\label{chiral10}
\ee

\subsubsection*{Non-perturbative effects}

Finally, an essential piece of the model are the non-perturbative effects whose source is located at a 4-cycle $S_{\rm np} \subset B$ whose embedding is defined by a holomorphic divisor function $h(x,y,z)$. As discussed in section \ref{s:review} such effects will shift the tree-level superpotential to (\ref{suponp0intro}), where $\theta_0 = (4\pi^2 m_*)^{-1} [{\rm log\, } h]_{z=0}$. As the specific value for $\theta_0$ depends on $S_{\rm np}$ and hence on the global completion of the local model, we will assume $\theta_0$ to be a general holomorphic function on $x$, $y$ that near the Yukawa point can be approximated by
\be
\theta_0\, =\, i (\theta_{00} + \theta_x x + \theta_y y)
\label{theta0}
\ee

In general, the presence of such non-perturbative effects will modify the local $E_6$ model described above, in the sense that the shift in the 7-brane superpotential modifies the F-term equations (\ref{Fterm7}) to
\begin{subequations}
\label{Fterm7np}
\begin{align}
\bar \partial_{A} \Phi +\eps \, \p\theta_0 \wedge F= &\, 0\\
F^{(0,2)} =&\, 0
\end{align}
\end{subequations}
and so $\langle F \rangle$ and $\langle \Phi \rangle$ need to be shifted from the original values in order to satisfy these new equations. These non-perturbative corrections to the 7-brane background for the $E_6$ model will be computed in subsection \ref{ss:zmnp}. Nevertheless, as shown in \cite{fimr12} such corrections to the background cancel each other out in the computation of holomorphic Yukawa couplings, and so one may still consider (\ref{sumPhi}) and (\ref{sumF}) for such purpose. Using this fact, in the next section we will show that the effect of (\ref{suponp0intro}) is to generate a hierarchical rank 3 matrix of up-type holomorphic Yukawa couplings.

\section{Holomorphic Yukawas via residues}
\label{s:holoyuk}

The purpose of this section is to compute the holomorphic piece of the $\mathbf{10} \times \mathbf{10} \times \mathbf{5}$ Yukawa couplings for the $E_6$ model above, and to show that the effect of the non-perturbative superpotential in (\ref{suponp0intro}) is to increase the rank of this Yukawa matrix from one to three. As pointed out in \cite{cchv09} holomorphic Yukawas in intersecting 7-brane models can be computed via an elegant residue formula that only depends on the 7-brane background data around the Yukawa point. Such residue formula was generalized to include the effect of the non-perturbative superpotential (\ref{suponpintro}) in \cite{fimr12}, and to include T-brane configurations in \cite{cchv10}. Our first task will then be to generalize all these previous results and derive a residue formula that includes both T-brane configurations and non-perturbative effects, mainly following the computations of Appendix D of \cite{fimr12}.

\subsection{Non-perturbative Yukawas and residues }

As explained in Section \ref{s:review}, in  order to compute 7-brane Yukawa couplings in the presence of non-perturbative effects we need to consider the superpotential
\be\label{eq:sup}
W=m_*^4 \int_S \tr(\Phi\, \w\, F)+\frac{\epsilon}{2}\theta_0\tr(F\, \w\, F)
\ee
where $\theta_0$ is a holomorphic section on $S$ and $\eps$ is a small parameter that measures the strength of the non-perturbative effects. From this superpotential follow the F-term equations (\ref{Fterm7np}) that together with the D-term equation (\ref{FI7}) form the equations of motion to be solved for the 7-brane background and zero modes. 

To proceed we separate the 7-brane bosonic fields as
\be
\Phi_{xy}\, =\, \langle \Phi_{xy} \rangle + \varphi_{xy} \quad \quad A_{\bar{m}}\, =\, \langle A_{\bar{m}} \rangle + a_{\bar{m}}
\label{fluct2}
\ee
and expand the equations of motion to linear order in the fluctuations $(\varphi, a)$ and their conjugate fields $(\varphi^\dag, a^\dag)$. From the F-terms (\ref{Fterm7np}) we obtain the zero mode equations
\begin{subequations}
\label{eq:flucF}
\begin{align}
\label{eq:fluc1}
\pa a&=0\\\label{eq:fluc3}
\pa\varphi-i[a,\lp]+\epsilon\partial\theta_0\,\w\,(\partial_{\langle A\rangle}a+\bar{\partial}_{\langle A\rangle}a^\dag)&=0
\end{align}
\end{subequations}
while the D-term gives
\be
\omega \wedge (\ph a + \pa a^\dag) - \frac{1}{2} \left( [\langle \bar{\Phi} \rangle, \varphi] + [\vphi^\dag,\langle \Phi \rangle]\right)\, =\, 0
\label{eq:fluc2}
\ee
Here $\langle A \rangle$ and $\langle \Phi \rangle$ are such that they satisfy the equations of motion (\ref{Fterm7np}) and (\ref{Dterm7}) at the level of the background. Using this fact and following \cite{fimr12} one obtains that the general solution to the F-term equations (\ref{eq:flucF}) is given by
\begin{subequations}
\label{genF}
\begin{align}
\label{gena}
a&=\pa \xi\\\label{genphi}
\varphi&=h-i[\lp,\xi]+\epsilon\partial\theta_0\,\w\,(a^\dag-\ph\xi)
\end{align}
\end{subequations}
where $\xi$ is a 0-form in the adjoint representation of complexified algebra (in our case $\mathfrak{e}_6^{\mathbb C}$), and $h$ is a $(2,0)$-form also in the adjoint and such that $\pa h=0$.

We may now consider again the superpotential (\ref{eq:sup}) and expand it to cubic order in fluctuations in order to compute the Yukawa couplings via the triple overlap of zero modes. Notice that the superpotential piece proportional to $\eps$ introduces a dependence on $a^\dag$ in such triple overlap. Nevertheless, when taking into account the solutions (\ref{genF}) one can show that all the terms containing the fluctuations $a^\dag$ arrange themselves into total derivatives and do not contribute to the Yukawa couplings. We refer the reader to Appendix D of \cite{fimr12} for a more detailed discussion of this point, and here we simply state the final result, namely that Yukawa couplings are computed from the integral
\be
Y \, =\, -i m_*^4 \int_S \tr \left(\varphi\, \w\, a\, \w\, a \right) 
\label{yukubic}
\ee
where the zero mode components $(a, \vphi)$ have the form
\begin{subequations}
\label{genF2}
\begin{align}
\label{gena}
a&=\pa \xi\\\label{genphi2}
\varphi&=h-i[\lp,\xi]- \epsilon\partial\theta_0\,\w\,\ph\xi
\end{align}
\end{subequations}
with $\xi$ and $h$ as above. Using these expressions (\ref{yukubic}) can be rewritten as
\begin{equation}
Y =  -i \frac{m_*^4}{3}\int_S \mathrm{Tr} \big( h \wedge a \wedge a -  \pa (\varphi \wedge [a ,\xi] )-\epsilon \ph (\theta_0 \ph (a \wedge a \,\xi))\big)
\label{interyuk}
\end{equation}
Since $\ph$, $\pa$ act on gauge invariant objects they reduce respectively to $\p$ and $\bar{\p}$, so the last two terms in (\ref{interyuk}) are boundary terms that vanish upon integration because they involve localized fields $(a, \vphi)$. The first term can be expressed as
\begin{equation}
Y =  -i \frac{m_*^4}{3} \int_S \mathrm{Tr} \left( h \wedge \pa \xi  \wedge \pa \xi \right)  
\label{yukhxixi}
\end{equation}
and can be computed by evaluating a residue at the Yukawa point. To see this it is convenient to use the invariance of the superpotential (\ref{yukubic}) under complexified gauge transformations
\be
\begin{array}{rcl}
a&\rightarrow& a+\bar{\partial}_{\langle A \rangle}\chi\\
\varphi&\rightarrow& \varphi-i\left [\lp,\chi\,\right ]
\end{array}
\label{cgt}
\ee
in order to take the 7-brane background to the case where $\langle A_{0,1} \rangle =0$, usually dubbed holomorphic gauge \cite{fi09,cchv10}. There the covariant derivative $\pa$ is replaced by $\bar{\partial}$ and the Higgs background reads
\be
\langle \Phi  \rangle \, = \, \langle\Phi \rangle^{(0)} +  \eps\,  \p \th_0 \wedge \langle A_{1,0}\rangle^{(0)}
\label{bkgeps}
\ee
where $\langle \Phi \rangle^{(0)}$,   $\langle A \rangle^{(0)}$ stand for the solution to the background equations of motion in the limit $\eps \raw 0$ and in the holomorphic gauge. As a result, $\xi$ satisfies the equation
\be
\Psi \xi\, dx \wedge dy\, = \, i (\varphi - h + \eps \p\theta_0 \wedge \p \xi)
\label{genphi3}
\ee
where $\Psi$ is a holomorphic matrix defined by $[\langle \Phi \rangle^{(0)}, \xi] = \Psi \xi\, dx \wedge dy$. This equation can be solved in perturbation theory, obtaining
\be
\begin{array}{c}
\xi  =  \xi^{(0)} +  i\eps\, \Psi^{-1} \left( \p_x\th_0\p_y \xi^{(0)} - \p_y\th_0\p_x \xi^{(0)} \right) + \CO(\eps^2)\\
\xi^{(0)}  =  i \Psi^{-1} \left(\varphi_{xy} - h_{xy} \right)
\end{array}
\label{solxi}
\ee
One may then plug this solution for $\xi$ into (\ref{yukhxixi}) with $\pa \raw \bar{\p}$ and, by integrating the total derivatives, convert it into a surface integral around the Yukawa point. As in \cite{cchv09,cchv10,fimr12},  the localized modes $\vphi_{xy}$ that appear in (\ref{solxi}) do not contribute, and we end up with an expression of the form 
\be
Y\, =\, m_*^4 f_{abc} \int_{\CR} (\eta^a\eta^bh_{xy}^c)\, dx \wedge dy 
\label{yukheta}
\ee
where $f_{abc}$ are structure constants of the symmetry group $G_p$ at the Yukawa point $p$, $\CR$ is diffeomorphic to the product of two circles surrounding $p$, and $\eta$ are the auxiliary holomorphic functions
\be
\eta\, =\, -i\Psi^{-1} h_{xy} + \eps \Psi^{-1} \left(\p_x\th_0\p_y (\Psi^{-1}h_{xy}) - \p_y\th_0\p_x (\Psi^{-1}h_{xy}) \right) + \CO(\eps^2)
\label{eta}
\ee
related to $\xi$ by removing the dependence on $\vphi_{xy}$. Finally, we can express (\ref{yukheta}) as a residue formula evaluated at the Yukawa point $p$
\be
Y\, = \, m_*^4  \pi^2 f_{abc}\, {\rm Res\, }_p (\eta^a\eta^bh^c)
\label{yukres}
\ee
where for simplicity we have removed the subindices to $h_{xy}$. In the following we will apply this residue formula to the $E_6$ local model constructed in Section \ref{s:E6}.

\subsection{Holomorphic Yukawas}

In order to apply the above residue formula to the $E_6$ model of Section \ref{s:E6} let us first gather the information which is relevant for computing the residue. Clearly, in order to compute the residue we only need to know the details of the model around the Yukawa point $p_{\rm up} =\{ x=y=0\}$, and so the local description of the $E_6$ model that was given in Section \ref{s:E6} is justified. Moreover, from all the parameters that are involved in the local $E_6$ model only a few of them are relevant for computing (\ref{yukres}). In fact, as can be deduced from our previous discussion there are basically only two quantities which are relevant in the computation of the residue: the Higgs background $\langle \Phi^{\rm hol} \rangle$ that solves the equations of motion in the holomorphic gauge and in the absence of non-perturbative effects, and the holomorphic function $\theta_0$ that encodes the information of such effects in the vicinity of the Yukawa point. For the reader's convenience we repeat both quantities here:
\bea
\label{Higgshol}
\langle \Phi_{xy}^{\rm hol} \rangle^{(0)} & =& m( E^+ + m\,x E^-) +\mu^2 (bx-y) Q\\
\theta_0 &  = & i (\theta_{00} + \theta_x x + \theta_y y)
\label{thetahol}
\eea
As discussed in section \ref{s:E6} the Higgs vev (\ref{Higgshol}) specifies the two matter curves $\Sigma_{\mathbf{5}}$ and $\Sigma_{\mathbf{10}}$ where the chiral modes of the $\mathbf{5}$-plets and $\mathbf{10}$-plets are localized. For each of these two sectors we need to specify the pair $(h, \eta)$ that will enter into the residue formula (\ref{yukres}), and will couple to each other via the structure constants $f_{abc}$ of $E_6$. 

\subsubsection*{Sector 5}

In this case the matter curve is given by $\Sigma_{\mathbf{5}} = \{bx-y=0\}$ and there the localized zero modes may arise in two possible sectors: along $E_{\mathbf{5}} = \oh ( \sqrt{3},\underline{1,-1,-1,-1,-1})$ and along $E_{\bar{\mathbf{5}}} = \oh (-\sqrt{3},\underline{-1,1,1,1,1})$. We will consider the case where, due to the presence of worldvolume fluxes, we have a chiral spectrum and a single zero mode in the sector $\mathbf{5}$ and none in $\bar{\mathbf{5}}$.\footnote{For vanishing hypercharge flux, and for the choice (\ref{choiceab}) the condition that chiral modes localized near the Yukawa point arise from the $\mathbf{5}$ sector is implemented by (\ref{cond23b}), while the fact that this is the only zero mode at this curve depends on the global aspects of the model, and we will take it as an assumption. When introducing the hypercharge flux $N_Y$ and imposing the condition (\ref{cond23}) the $SU(5)$ spectrum will be broken and there will only be a localized mode in the sector $\mathbf{5}_1$, namely the MSSM Higgs doublet $H_u$. The holomorphic Yukawas computed in this section will also be valid for the case, with the only replacement $\mathbf{5} \raw \mathbf{5}_1$. See sections \ref{s:zeromodes} and \ref{s:physyuk} for more details.} The action of (\ref{Higgshol}) in this zero mode sector is such that
\be
\Psi\, =\, 2\mu^2 (bx-y)
\ee
It then only remains to specify the value of $h$ for this sector. While in principle $h= h(x,y)$ may be any holomorphic function in the vicinity of $x=y=0$, one may follow the philosophy in \cite{cchv09} and apply a gauge transformation of the form (\ref{cgt}) with $\chi$ holomorphic. Such transformation will not take us away from the holomorphic gauge and will be able to remove any dependence of $h$ on the complex coordinate $bx-y$. We then have that in this sector $h$ can be taken to be an arbitrary holomorphic function of the orthogonal coordinate $x+by$. Because by assumption we only have one zero mode we will take it to be a constant, following the standard practice in the literature \cite{hv08}. We then have that
\bea
\label{h5}
h_{\mathbf{5}}/\g_{\mathbf{5}} & = & 1\\
i \eta_{\mathbf{5}}/\g_{\mathbf{5}} & = &  \frac{1}{ 2\mu^2 (bx-y)} - \eps \frac{\th_{x} + b \th_{y}}{4\mu^4(bx-y)^3} + \CO(\eps^2)
\eea
with $\g_{\mathbf{5}}$ a real constant to be computed via wavefunction normalization in the next section. 

\subsubsection*{Sector 10}

In this case the curve is given by $\Sigma_{\mathbf{10}} = \{\mu^4 (bx-y)^2 = m^3 x\}$ and the localized modes live in the root subspace spanned by (\ref{10roots}). As before, we will assume that worldvolume fluxes are such there are exactly three chiral zero modes within the subspace spanned by $E_{\mathbf{10}^+} = (0,\underline{1,1,0,0,0})$ and $E_{\mathbf{10}^-} = \oh (-\sqrt{3},\underline{1,1,-1,-1,-1})$ (see Appendix \ref{ap:chiral} for more details) and that these will be our three families of $\mathbf{10}$-plets in our $SU(5)$ GUT model. 

The action of (\ref{Higgshol}) in this sector is such that
\be
\Psi
\left(
\begin{array}{c}
E_{\mathbf {10}^+} \\ E_{\mathbf {10}^-}
\end{array}
\right)
\, =\, 
\left(
\begin{array}{cc}
-\mu^2(bx-y) & m \\ m^2 x & -\mu^2(bx-y)
\end{array}
\right)
\left(
\begin{array}{c}
E_{\mathbf {10}^+} \\ E_{\mathbf {10}^-}
\end{array}
\right)
\label{cocacola}
\ee
as can be read from (\ref{com10}). As before we need to specify $h_{\mathbf{10}}$, which now will be an $SU(2)$ doublet of arbitrary holomorphic functions. Again, by performing an appropriate holomorphic gauge transformation (\ref{cgt}) we can restrict ourselves to a very particular form for $h_{\mathbf{10}}$ since \cite{cchv10}
\be
h_{\mathbf{10}}\, =\, 
\left(
\begin{array}{c}
h^+(x,y) \\ h^-(x,y)
\end{array}
\right)
- i \Psi 
\left(
\begin{array}{c}
\chi^+(x,y) \\ \chi^-(x,y)
\end{array}
\right)\, =\, 
\left(
\begin{array}{c}
0\\ h(bx-y)
\end{array}
\right)
\ee
for arbitrary $h^\pm$ and appropriate choices of $\chi^\pm$. While $h$ can be any holomorphic function on the coordinate $bx-y$, under the assumption that we have three zero modes in this sector we can take them to be the monomials $\g_{\mathbf{10}}^i m_*^{3-i} (bx-y)^{3-i}$, with $\g_{\mathbf{10}}^i$ some normalization factors to be fixed in the next section. We finally have that
\bea
\label{solh10}
h_{\mathbf{10}}^i/\g_{\mathbf{10}}^i & = &  \left(\begin{array}{c} 0 \\ m_*^{3-i} (bx-y)^{3-i} \end{array}\right) \\[2mm] \label{soleta10}
i \eta_{\mathbf{10}}^i/\g_{\mathbf{10}}^i & = & - \left[ \frac{m_*^{3-i} (bx-y)^{3-i}  }{ \mu^4 (bx-y)^2 - m^3 x}\right]   \left(\begin{array}{c}m\\\mu^2 (bx-y)  \end{array}\right) + \CO(\eps^2) \\ \nonumber
& + & \eps\,   \frac{2\mu^4 (\th_x+b\th_y)(bx-y) + m^3\th_y}{(\mu^4 (bx-y)^2 - m^3 x)^3} m_*^{3-i} (bx-y)^{3-i}\left(\begin{array}{c} 2 m \mu^2 (bx-y) \\(m^3x+\mu^4(bx-y)^2)\end{array}\right)\\ \nonumber
& + &  \eps\, \frac{ (\th_x+b\th_y)}{(\mu^4 (bx-y)^2 - m^3 x)^2}  m_*^{3-i} (bx-y)^{2-i}  \left(\begin{array}{c}2m \mu^2 (bx-y)(6-i)\\m^3x(3-i)+(4-i)\mu^4(bx-y)^2\end{array}\right)
\eea
which has a rather complicated $\CO(\eps)$ correction to $\eta_{\mathbf{10}}^i$. Nevertheless, the result that one obtains from applying the residue formula is still quite simple, as we will now see.

\subsubsection*{$\mathbf{10 \times 10 \times 5}$ Yukawas}

Let us now apply the explicit expressions for $(h_{\mathbf{5}}, \eta_{\mathbf{5}})$ and $(h_{\mathbf{10}}, \eta_{\mathbf{10}})$ to the residue formula (\ref{yukres}) for the Yukawa couplings. An important simplifications arises from the fact that the structure constants of $E_6$ satisfy
\begin{equation}\label{e6comm}
\mathrm{Tr}([E_{\mathbf{5}\, i}, E_{\mathbf{10}\, jk}^{\quad M}] E_{\mathbf{10}\, lm}^{\quad N}) = \epsilon_{ijklm}\epsilon^{MN}
\end{equation}
where $i,j,k,l,m$ are $\mathfrak{su}(5)$ indices and $M,N =\pm$ are $\mathfrak{su}(2)$ indices. As a result the non-trivial contributions to the $\mathbf{10 \times 10 \times 5}$ Yukawa will be of the form
\be
Y\, =\, m_*^4  \pi^2 {\rm Res\, }_{(0,0)} \left(\eps_{MN} \eta_{\mathbf{5}}\eta_{\mathbf{10}}^M h_{\mathbf{10}}^N \right)\, =\, m_*  \pi^2 {\rm Res\, }_{(0,0)} \left(\eta_{\mathbf{5}}\eta_{\mathbf{10}}^+ h_{\mathbf{10}}^- \right)
\label{resu5}
\ee
where the contractions of the $SU(5)$ indices have been left implicit. In the first equality we have used that any other contribution will contain a term of the form $\eps_{MN} \eta_{\mathbf{10}}^M \eta_{\mathbf{10}}^N$ and so it will vanish identically, and in the second equality we have used that in our solution (\ref{solh10}) $h_{\mathbf{10}}^+ = 0$. Hence, even if (\ref{soleta10}) has a complicated expression only the terms proportional to $E_{\mathbf{10}^+}$ will be relevant when computing up-like Yukawa couplings. 

Let us proceed by computing (\ref{resu5}) explicitly. At zeroth order in $\eps$ we have a contribution of the form 
\bea
\label{yukholtree}
Y^{ij}_{\rm tree} & = & m_*^4 \pi^2 \g_{\mathbf{5}} \g_{\mathbf{10}}^i \g_{\mathbf{10}}^j\, {\rm Res\, }_{(0,0)} \left[ \frac{m (m_* (bx-y))^{6-i-j}}{2\mu^2(bx-y)(\mu^4(bx-y)^2 -  m^3 x)} \right] \\ \nonumber
& = & - \frac{m_*^4 \pi^2}{2m^{2} \mu^{2}} \g_{\mathbf{5}} \g_{\mathbf{10}}^i \g_{\mathbf{10}}^j\, \d_{i3}\d_{j3}
\eea
and so at this level only $Y^{33}$ is non-zero.
At order $\CO(\eps)$ we get a contribution of the form
\be
Y^{ij}_{\rm np}\, =\, \eps\,\frac{m_*^6 \pi^2}{4m^{2} \mu^{4}} \left[ b\th_y + \th_x \right] \g_{\mathbf{5}} \g_{\mathbf{10}}^i \g_{\mathbf{10}}^j\, \d_{(i+j)4}
\label{yukholnp}
\ee
from the $\CO(\eps)$ correction to $\eta_{\mathbf{5}}$. In fact, one can check that the $\CO(\eps)$ correction to $\eta_{\mathbf{10}}$ do not contribute to (\ref{resu5}) and that we are left with the following $\mathbf{10 \times 10 \times 5}$ Yukawa couplings:
\be
Y^{ij} = \frac{\pi^2\g_{\mathbf{5}}}{4\rho_\mu\rho_m}
\left(
\begin{array}{ccc}0&0& \tilde \eps \rho_\mu^{-1} \g_{\mathbf{10}}^1\g_{\mathbf{10}}^3 \\0& \ \tilde \eps \rho_\mu^{-1} \g_{\mathbf{10}}^2\g_{\mathbf{10}}^2 &0\\ \tilde \eps \rho_\mu^{-1} \g_{\mathbf{10}}^1\g_{\mathbf{10}}^3 &0&- 2 \gamma_{\mathbf{10}}^3 \g_{\mathbf{10}}^3
\end{array}
\right)
+\mathcal{O}(\eps^2)
\label{Yukhol}
\ee
where we have defined the slope densities
\be
\rho_{\mu} \, =\, \frac{\mu^2}{m_*^2} \qquad \qquad \rho_m  \, =\, \frac{m^2}{m_*^2} 
\label{rhos}
\ee
as well as the non-perturbative parameter
\be
\tilde \eps\, =\, \eps\, (\th_x  + b\th_y)
\ee

As claimed, we obtain a Yukawa matrix such that in the absence of non-perturbative effects has rank one, but when taking them into account increases its rank to three.\footnote{More precisely, the condition for rank enhancement is that $\tilde \eps \neq 0$, which seems to indicate that the pull-back of $\th_0$ along $\Sigma_{\mathbf{5}}$ must be non-trivial.} Note that the eigenvalues of this matrix display a hierarchical structure $(\CO(1), \CO(\tilde\eps), \CO(\tilde\eps^2))$, as we will discuss in more detail in Section \ref{s:physyuk}. 

An interesting feature of this Yukawa matrix it that its entries depend on very few parameters of the model, most notably $\tilde{\eps}$, $\rho_\mu$ and $\g_{\mathbf{10}}^i$. In fact the last set of parameters can be understood as wavefunction normalization constants that cannot be determined from the analysis of this section. Instead, they can be calculated by computing zero mode fluctuations in a physical background and demanding that their 4d kinetic terms are canonically normalized, which is the task that we will endeavor in the next section. As we will see, $\g_{\mathbf{10}}^i$ will depend on the worldvolume flux densities of the model, and in particular in the hypercharge flux densities in (\ref{Yflux}). As U-quarks with different hypercharge feel $F_Y$ differently, $\g_{\mathbf{10}}^i$ will take different values for each of them, and this will give rise to a rich structure of physical Yukawa couplings to be analyzed in Section \ref{s:physyuk}.

\section{Zero mode wavefunctions at the $E_6$ point}
\label{s:zeromodes}

An remarkable aspect of the computations of the last section is that, in order to arrive to the Yukawa matrix (\ref{Yukhol}), we did not have to fully solve for the chiral zero mode wavefunctions. Instead, we solved for the F-term equations and used the invariance of the superpotential under complexified gauge transformations. The price to pay for using that trick is that we do not have any physical criterium to fix the constants $\g_{\mathbf{5}}$, $\g_{\mathbf{10}}^i$ that appear in the Yukawa matrix, because the wavefunctions that we are using are not in a physical gauge. As pointed out in \cite{cchv09} this is because via our previous computation we are only computing the holomorphic piece of the Yukawa couplings, and not their actual physical values. In order to compute physical Yukawa couplings we also need to solve the D-term equations for the zero mode wavefunctions and the demand that their corresponding 4d fluctuations have canonically normalized kinetic terms. This will fix the constants $\g_{\mathbf{5}}$, $\g_{\mathbf{10}}^i$ in terms of the data of the local model and provide us with the physical Yukawa matrix to be analyzed in the next section. 

As we will see, solving analytically for the zero modes D-term equations is a rather involved task, mainly because they involve the Painlev\'e transcendent $f$ found in subsection \ref{ss:Tbranebkg}. Nevertheless, we will be able to do so for a certain region of parameters of our local model, and we expect that our general conclusions are valid for other regions as well. We will first compute these physical wavefunctions in the absence of non-perturbative effects, which will already allow us to compute the normalization factors $\g_{\mathbf{5}}$, $\g_{\mathbf{10}}^i$ to a good approximation. We will then include the corrections induced by non-perturbative, in the spirit of \cite{afim,fimr12}. As a cross-check of our results, we will use the corrected wavefunctions to rederive the Yukawa matrix (\ref{Yukhol}), now with the factors $\g_{\mathbf{5}}$, $\g_{\mathbf{10}}^i$ fixed.

\subsection{Perturbative zero-modes}
\label{ss:zmp}

In the absence of non-perturbative effects (i.e., $\eps=0$) the zero mode equations (\ref{eq:flucF}) and (\ref{eq:fluc2}) reduce to
\bea
\label{treeF1}
\pa a&=&0\\
\label{treeF2}
\pa\varphi+i[\lp,a]& = & 0\\
\label{treeD}
\omega \wedge \ph a - \frac{1}{2} [\langle \bar{\Phi} \rangle, \varphi] & = & 0
\eea
In fact, while the above equations are written for bosonic fluctuations, the same equations apply for the 7-brane fermionic zero modes, pairing up into 4d $\cn=1$ chiral multiplets $(a_{\bar{m}}, \psi_{\bar{m}})$ and $(\varphi_{xy}, \chi_{xy})$ with the same internal profile. In the following we will display the solutions to these equations for both the $\mathbf{5}$ and $\mathbf{10}$ sectors of the $E_6$ model, leaving most of the technical computations to Appendix \ref{ap:wave}.

\subsubsection*{Sector 5}

To solve for this sector it is useful to write the Ansatz
\be
\left(
\begin{array}{c}
a_{\bar{x}}\\ a_{\bar{y} }\\ \vphi_{xy}
\end{array}
\right)
\, =\, 
\vec{\vphi}_{\mathbf{5}} E_{\mathbf{5}} \qquad \qquad  E_{\mathbf{5}}\, =\,  \oh ( \sqrt{3},\underline{1,-1,-1,-1,-1})
\ee
so that in a particular gauge for $\langle A \rangle $ the zero mode equations translate into
\be
\left(
\begin{array}{cccc}
0 & D_x & D_y & D_z \\
-D_x & 0 & -D_{\bar{z}} & D_{\bar{y}} \\
-D_y & D_{\bar{z}} & 0 & -D_{\bar{x}} \\
-D_z & -D_{\bar{y}} & D_{\bar{x}} & 0
\end{array}
\right)
\left(
\begin{array}{c}
0 \\ \\  \vec{\vphi}_{\mathbf{5}} \\ \quad
\end{array}
\right)\, =\, 0  
\label{Dirac5}
\ee
with
\be
D_{x} \, =\, \p_{x} + \oh(q_R \bar{x} - q_S \bar{y}) \qquad D_{y} \, =\, \p_{y} - \oh (q_R \bar{y}  + q_S \bar{x}) \qquad D_{z}\, =\, 2i \mu^2 (\bar{x}-\bar{y}) 
\label{covar}
\ee
and $D_{\bar{m}}$ their conjugates. The quantities $q_R$ and $q_S$ are constants the depend on the flux densities of the model as indicated in table \ref{t:sectors}, and for concreteness we have taken the choice (\ref{choiceab}) for the Higgs background $\langle \Phi_{xy}\rangle$. 

Following \cite{fimr12} one can easily solve this system of equations, obtaining 
\be
\label{wave5p}
\vec\vphi_{\mathbf{5}}\, =\, \gamma_{\mathbf{5}}\left ( \begin{array}{c}
i\frac{\zeta_{\mathbf{5}}}{2\mu^2}\\
i\frac{(\zeta_{\mathbf{5}}-\lam_{\mathbf{5}})}{2\mu^2}\\
1\end{array}\right )\, \chi_{\mathbf{5}}, \quad
\qquad
\chi_{\mathbf{5}} = 
e^{\frac{q_R}{2}(|x|^2-|y|^2)-q_S (x \bar y +y\bar x)+(x-y)(\zeta_{\mathbf{5}}\bar x-(\lam_{\mathbf{5}}-\zeta_{\mathbf{5}})\bar y))}
\ee
with $\lam_{\mathbf{5}}$ the lowest solution to 
\be
\lam^3_{\mathbf{5}}-(8\mu^4+(q_R)^2+(q_S)^2)\lam_{\mathbf{5}}+8\mu^4q_S=0
\ee
and $\zeta_{\mathbf{5}}=\frac{\lam_{\mathbf{5}}(\lam_{\mathbf{5}}-q_R-q_S)}{2(\lam_{\mathbf{5}}-q_S)}$, see Appendix \ref{ap:wave} for further details.\footnote{As shown in the appendix we may multiply $\chi_{\mathbf{5}}$ by an arbitrary holomorphic function of a linear combination of $x$ and $y$ and find further solutions to the zero mode equations. By assumption there should be a single zero mode in this sector, hence a single holomorphic function specified by the global geometry of the model. Nevertheless the main contribution to the Yukawa couplings comes from the average value of such function around the Yukawa point, so we may safely approximate it by a constant, consistently with the choice made in eq.(\ref{h5}).}

Notice that, because they depend on the hypercharge flux, $q_R$ and $q_S$ take different values for the two subsectors $\mathbf{5}_1$ and $\mathbf{5}_2$ of table \ref{t:sectors}, and so the same is true for $\lam_{\mathbf{5}}$, $\zeta_{\mathbf{5}}$. In particular, imposing (\ref{cond23}) we find that $q_S(\mathbf{5}_2)=0$ and that the wavefunction for this sector is not localized along $\Sigma_{\mathbf{5}}$, as we briefly comment below. 

\subsubsection*{Sector 10}

This sector is more involved because the zero modes lie along the root subspace spanned by $E_{\mathbf{10}^+} = (0,\underline{1,1,0,0,0})$ and $E_{\mathbf{10}^-} = \oh (-\sqrt{3},\underline{1,1,-1,-1,-1})$ and so the appropriate Ansatz is
\be
\left(
\begin{array}{c}
a_{\bar{x}}\\ a_{\bar{y} }\\ \vphi_{xy}
\end{array}
\right)
\, =\, 
\vec{\vphi}_{\mathbf{10}^+} E_{\mathbf{10}^+} + \vec{\vphi}_{\mathbf{10}^-} E_{\mathbf{10}^-}
\ee
from which one can write an equation analogous to (\ref{Dirac5}). Because $E_{\mathbf{10}^\pm}$ transform as a doublet of the $SU(2)$  generated by $\{E^+,E^-,P\}$, c.f.(\ref{Edoublet}), it is useful to represent these wavefunction components with the following doublet notation
\be
a = \bmat{c} a^+ \\
a^-
\emat \qquad  \qquad
\vphi = \bmat{c} \vphi^+ \\
\vphi^-
\emat
\label{doubs}
\ee
where $a_{\bar{x}}^\pm$, $a_{\bar{y}}^\pm$, $\vphi_{xy}^\pm$ belong respectively to $\vec{\varphi}_{\mathbf{10}^\pm}$. Then, following the strategy in \cite{afim,fimr12}, we use the solution for the F-terms equations (\ref{treeF1}) and (\ref{treeF2}) to write $a$ in terms of $\varphi$, and then substitute in the D-term equation (\ref{treeD}) to find an equation for $\vphi$. 

It is instructive to first consider the case where the primitive flux $\langle F_p \rangle$ in (\ref{sumFp}) is absent. Then solution to the F-term equations is in fact quite similar to the one found in the previous section in the holomorphic gauge, c.f. (\ref{genF2}), and reads
\begin{subequations}
\label{Fsol}
\begin{align}
\label{asol}
a & =  e^{fP/2} \bar\p \xi\\
\label{phisol}
\vphi & =  e^{fP/2}\left(h - i \Psi \xi \right) 
\end{align}
\end{subequations} 
where $\xi$ and $h$ are also doublets with components $\xi^\pm$ and $h^\pm$ and 
\be
P = \bmat{cc} 1 & 0 \\
0 & -1
\emat \qquad  \qquad
\Psi = \bmat{cc} -\mu^2(x-y) & m \\
m^2 x & -\mu^2(x-y)
\emat
\label{pepsi}
\ee
In particular, notice that $\Psi$ is the same matrix as in (\ref{cocacola}) after taking the choice (\ref{choiceab}). From (\ref{phisol}) we obtain
\be
\xi = i \Psi^{-1} \left( e^{-fP/2}\vphi - h\right)
\label{xisol}
\ee
which is the analogue of the lower equation in (\ref{solxi}) for the physical background (\ref{sumPhi}). Finally, the D-term equation for the fluctuations (\ref{treeD}) reads
\be
\p_x a_{\bar x} + \p_y a_{\bar y} + \frac12 \p_x f P a_{\bar x} - i e^{-fP/2}\Psi^\dagger e^{fP/2}\vphi = 0   
\label{dterm1} 
\ee 
which by using (\ref{asol}), and recalling that $f$ only depends on $x, \bar{x}$, we find 
\be
\p_x \p_{\bar x} \xi + \p_y \p_{\bar y} \xi + \p_x f P \p_{\bar x} \xi - i \Lambda^\dagger \left(h - i \Psi \xi \right)  = 0   
\label{dterm2} 
\ee 
where we have defined
\be
\Lambda =  e^{fP} \Psi  e^{-fP} = 
\bmat{cc} -\mu^2(x -y) & m e^{2f} \\
m^2 x e^{-2f} & -\mu^2(x-y)
\emat
\label{ladef}
\ee
To proceed it is convenient to make the following change of variables
\be 
U =  e^{-fP/2}\vphi  \qquad \Rightarrow \qquad \xi \,=\, i \Psi^{-1} \left(U - h\right)
\label{xisol2} 
\ee
and express (\ref{dterm2}) entirely in terms of the doublet $U$
\be
\p_x \p_{\bar x} U + \p_y \p_{\bar y} U 
-  (\p_x \Psi) \Psi^{-1}  \p_{\bar x} U  + (\p_y \Psi) \Psi^{-1} \p_{\bar y} U
+  \p_x f \Psi P \Psi^{-1} \p_{\bar x} U -  \Psi\Lambda^\dagger U = 0  
\label{uterm}
\ee
so that the dependence on $h$ drops completely. However, the D-term equation gives a coupled system of equations for $U^+$ and $U^-$ that are quite involved to solve. Nevertheless, as discussed in Appendix \ref{ap:wave} in the limit $m \gg \mu$ they decouple and one can prove that there is no localized mode for $U^+$, which we henceforth set to zero. Moreover, near the Yukawa point $p_{\rm up} =\{x=y=0\}$ one can approximate $f = \log c+  c^2 m^2 x\bar{x} + \dots$ and solve analytically for $U^-$, finding $U^- = {\rm exp} (\lam_{\mathbf{10}} x \bar x) h$ with $\lam_{\mathbf{10}}$ the negative solution to $ c^2 \lam_{\mathbf{10}}^3 + 4c^4 m^2 \lam_{\mathbf{10}}^2 - m^4\lam_{\mathbf{10}}=0$. At the end one finds the solution 
\be
\vec{\vphi}_{\mathbf{10}^+}^j \, =\, \gamma_{\mathbf{10}}^j
\left (\begin{array}{c}
\frac{i\lam_{\mathbf{10}}}{m^2}\\
0 \\
0 \end{array}\right )
 e^{f/2} \chi_{\mathbf{10}}^j 
 \qquad \quad 
\vec{\vphi}_{\mathbf{10}^-}^j \, =\, \gamma_{\mathbf{10}}^j
\left (\begin{array}{c}
0 \\
0 \\
1 \end{array}\right )
e^{-f/2} \chi_{\mathbf{10}}^j 
\ee
where $e^{f/2} = \sqrt{c}\, e^{m^2c^2x\bar{x}/2}$ and $\chi_{\mathbf{10}}^j \, =\, e^{\lam_{\mathbf{10}} x \bar x}\, g_j (y)$, with  $g_j$ holomorphic functions of $y$. 

Switching on the primitive worldvolume fluxes will amount to replace $\p_{x,y} \raw D_{x,y}$ in the D-term equation, with $D_{x,y}$ defined in (\ref{covar}), and similarly for $\bar{\p}$ in the F-term equations. Still, in the limit $m \gg \mu$ and near the origin one finds a localized solution for $U^-$ and the wavefunctions read
\be
\label{phys10} 
\vec{\vphi}_{\mathbf{10}^+}^j \, =\, \gamma_{\mathbf{10}}^j
\left (\begin{array}{c}
\frac{i\lam_{\mathbf{10}}}{m^2}\\
-\frac{i\lam_{\mathbf{10}}\zeta_{\mathbf{10}}}{m^2} \\
0 \end{array}\right )
 e^{f/2} \chi_{\mathbf{10}}^j 
 \qquad \quad 
\vec{\vphi}_{\mathbf{\mathbf{10}}^-}^j \, =\, \gamma_{\mathbf{10}}^j
\left (\begin{array}{c}
0 \\
0 \\
1 \end{array}\right )
e^{-f/2} \chi_{\mathbf{\mathbf{10}}}^j 
\ee
where $\lam_{\mathbf{10}}$ is the negative solution to 
\be
m^4 (\lambda_{\mathbf{10}} -q_R)+\lambda c^2 \left( c^2 m^2 (q_R-\lambda_{\mathbf{10}} )-\lambda_{\mathbf{10}} ^2+q_R^2+q_S^2\right)=0
\ee
and $\zeta_{\mathbf{10}} = -q_S/(\lam_{\mathbf{10}}-q_R)$. The scalar wavefunctions $\chi_{\mathbf{10}}$ read
\be
\chi_{\mathbf{10}}^j \, =\,e^{\frac{q_R}{2}(|x|^2-|y|^2)-q_S (x \bar y +y\bar x)+\lam_{\mathbf{10}} x (\bar x - \zeta_{\mathbf{10}} \bar y)} \, g_j (y +\zeta_{\mathbf{10}}x)
\label{wave10p}
\ee
where $g_j$ holomorphic functions of $y +\zeta_{\mathbf{10}}x$, and $j=1,2,3$ label the different zero mode families. Following \cite{hv08} we will choose such holomorphic representatives to be
\be
g_j\, =\, m_*^{3-j}(y +\zeta_{\mathbf{10}}x)^{3-j}
\label{holorep}
\ee
Finally, notice that within each family the wavefunctions differ for each of the sectors $\mathbf{10}_{1,2,3}$ of table \ref{t:sectors} because they have different hypercharges and so $q_R$ and $q_S$ take different values for each.  From the results of the previous sections we expect that this difference will only appear in the physical Yukawa couplings via different normalization factors $\g_{\mathbf{10}}^j$, which we now proceed to discuss.

\subsubsection*{Normalization factors}

Having obtained explicit expressions for the zero mode wavefunctions one may now require that the 4d chiral modes have canonically normalized kinetic terms. The 4d kinetic terms for the wavefunctions of a sector $\rho$ that one obtains via dimensional reduction are
\be
K_\rho^{ij}\, =\, \langle \vec{\vphi}_{\rho}^{i} | \vec{\vphi}_{\rho}^{j} \rangle \, = \,
m_*^2 \int_S \tr \,( \vec{\vphi}_{\rho}^{i}{}^\dag \cdot \vec{\vphi}_{\rho}^{j})\, {\rm d vol}_S
\label{4dnorm}
\ee
where $i$, $j$ are family indices. To have canonically normalized kinetic terms we need to impose that $K^{ij}_\rho = \d^{ij}$. In the case of the sector $\rho = \mathbf{5}$ there is only one family and we can easily achieve canonical kinetic terms by adjusting the value of the constant $\g_{\mathbf{5}}$. In this case the integral (\ref{4dnorm}) reads
\be
K_{\mathbf{5}}\, =\, m_*^2 |\g_{\mathbf{5}}|^2 ||\vec{v}_{\mathbf{5}}|| \, \int_S \chi_{\mathbf{5}}^* \chi_{\mathbf{5}}\, d{\text{vol}}_S
\label{4dnorm5}
\ee
with $\chi_{\mathbf{5}}$ given by (\ref{wave5p}), and $\vec{v}_{\mathbf{5}} = \frac{1}{2\mu^2} (i\zeta_{\mathbf{5}}, \, i(\zeta_{\mathbf{5}} - \lam_{\mathbf{5}}),\, 2\mu^2)^t$. Due to the convergence properties of $\chi_{\mathbf{5}}$ we can compute the above integral by extending the patch in which we define our local model to $\IC^2$. We find that the required value for $\g_{\mathbf{5}}$ is
\be
|\gamma_{\mathbf{5}}|^2=-\frac{4}{\pi^2}\left (\frac{\mu}{m_*} \right )^4\frac{(2\zeta_{\mathbf{5}}+q_R)(q_R+2\zeta_{\mathbf{5}}-2\lam_{\mathbf{5}})+(q_S+\lam_{\mathbf{5}})^2}{4\mu^4+\zeta_{\mathbf{5}}^2+(\zeta_{\mathbf{5}}-\lam_{\mathbf{5}})^2}
\label{norm5}
\ee
see Appendix \ref{ap:wave} of \cite{fimr12} for details of the derivation. Here $\lam_{\mathbf{5}}$ and $\zeta_{\mathbf{5}}$ are defined as in (\ref{wave5p}) and so depend on the worldvolume flux densities $q_R$ and $q_S$, which are given in table \ref{t:sectors} for both sectors $\mathbf{5}_1$ and $\mathbf{5}_2$. Hence in general both members of the $\mathbf{5}$-plet have different normalization factors. In fact, for the sector $\mathbf{5}_2 = (\mathbf{3}, \mathbf{1})_{1/3}$ that could contain a Higgs triplet we find that $\gamma_{5_2} = 0$ after we impose the condition (\ref{cond23}).\footnote{For $q_S=0$ the parameter $\zeta_{\mathbf{5}}$ defined below eq.(\ref{wave5p}) reduces to $\zeta_{\mathbf{5}}=\frac{1}{2}(\lam_{\mathbf{5}}-q_R)$ which upon substituting in (\ref{norm5}) shows that $\gamma_{{\mathbf{5}}_2}=0$.} That is because the integrand in (\ref{4dnorm}) is not localized along the curve $\{x=y\} \subset \IC^2$, which is turn related to the fact that this is a non-chiral sector of the model and one may assume that it only contains massive modes. 

Notice that for the $\mathbf{10}$ sector (\ref{4dnorm}) reads
\bea
K^{ij}_{\mathbf{10}} & = & m_*^2\int_{S}\tr(\vec{\vphi}_{\mathbf{10}^+}^i{}^\dagger\vec{\vphi}_{\mathbf{10}^+}^j+\vec{\vphi}_{\mathbf{10}^-}^i{}^\dagger\vec{\vphi}_{\mathbf{10}^-}^j)d\text{vol}_{S} \\ \nonumber
& = & m_*^2 (\g_{\mathbf{10}}^i)^*\g_{\mathbf{10}}^j \sum_{\kappa = \pm} ||\vec{v}_{\mathbf{10}^\kappa}|| \, \int_S e^{\kappa f} (\chi_{\mathbf{10}}^i)^* \chi_{\mathbf{10}}^j\, d{\text{vol}}_S
\label{4dnorm10}
\eea
with the vectors $\vec{v}_{\mathbf{10}^\pm}$ defined in (\ref{vec10}). Because the integrand needs to be invariant under the rotation $(x,y) \raw e^{i\a}(x,y)$ to have a non-vanishing result we deduce that $K^{ij}_{\mathbf{10}} = 0$ for $i \neq j$, and so we only need to adjust the constants $\g_{\mathbf{10}}^j$ in order to have canonical kinetic terms in this sector. In particular we obtain that the required result is
\be
|\gamma_{\mathbf{10}}^j|^2=-\frac{c}{m_*^2\pi^2(3-j)!}\frac{1}{\frac{1}{2\lam_{\mathbf{10}}+q_R(1+\zeta_{\mathbf{10}}^2)-m^2 c^2}+\frac{c^2\lam_{\mathbf{10}}^2}{m^4}\frac{1}{2\lam_{\mathbf{10}}+q_R(1+\zeta_{\mathbf{10}}^2)+m^2 c^2}}\left (\frac{q_R}{m_*^2}\right )^{4-j}
\label{norm10}
\ee
which not only depend on the family index $j$, but also on the sectors $\mathbf{10}_{1,2,3}$ of table \ref{t:sectors}, again via the flux densities $q_R$ and $q_S$ and the quantities $\lam_{\mathbf{10}}$, $\zeta_{\mathbf{10}}$ that depend on them. Finally, notice that the effects of the non-primitive flux (\ref{sumFp}) in this sector appear through the dependence on the constant $c$.

\subsection{Non-perturbative corrections}
\label{ss:zmnp}

Let us now see how the presence of non-perturbative effects modifies the above wavefunction profile. As stated before, at the level of approximation that we are working these effects amount to add the term proportional to $\eps$ in the F-term equation (\ref{Fterm7np}). This will modify the 7-brane background $\langle \Phi \rangle$ and $\langle A \rangle$ as well as the wavefunction profiles that were just computed for $\eps=0$. These deformations are particularly involved for the T-brane sector of our background and as a consequence for the wavefunctions of the $\mathbf{10}$ matter curve. Nevertheless, as we will see the $\CO(\eps)$ corrections to the $\mathbf{10}$-plet wavefunctions only affect the Yukawa couplings at $\CO(\eps^2)$, and so they can be neglected to the level of approximation of (\ref{Yukhol}). In the next section we will see that we can reproduce (\ref{Yukhol}) via the triple overlap of the $\CO(\eps)$ corrected wavefunctions, now with explicit expressions for the normalization factors $\g_{\mathbf{5}}$, $\g_{\mathbf{10}}^i$. 

\subsubsection*{Corrections to the background}

Following Section \ref{s:holoyuk}, we can solve the equations of motion for the background for $\eps\neq 0$ in the holomorphic gauge if we take $\langle A_{0,1} \rangle =0$ and $\langle \Phi  \rangle$ as in (\ref{bkgeps}). There $\langle\Phi \rangle^{(0)}$, $\langle A _{1,0}\rangle^{(0)}$ are given by the background at $\eps=0$ and in the holomorphic gauge. Let us first assume that the primitive fluxes (\ref{sumFp}) vanish. Then we have that $\langle \Phi \rangle^{(0)}$ is given by (\ref{TbranePhi}) and $\langle A _{1,0}\rangle^{(0)} = i \p f \, P$ and so in the holomorphic gauge
\be
\langle \Phi_{xy} \rangle\, =\, m( E^+ + m\,x E^-) + \eps \, \th_y \p_x f\, P + \mu^2 (x - y) Q
\ee
where we have used that $f = f(x, \bar{x})$ and taken the choice (\ref{choiceab}). One may now perform a complexified gauge transformation (\ref{cgtbkg}) in order to go to a real gauge that satisfies the D-term (\ref{Dterm7}) up to $\CO(\eps^2)$. For this we need generalize the Ansatz (\ref{eq:ans}) to
\be
g\, =\, e^{\frac{f}{2} P}e^{ \frac{\eps}{2} ({k} E^++ {k}^* E^-)}\, =\, e^{\frac{f}{2} P} + \frac{\eps}{2}({k\, }e^{f/2}E^++ {k}^*e^{-f/2}E^-) + \CO(\eps^2)
\label{cgtnp}
\ee
with $f$ as above and $k$ a complex function of $x,\bar{x}$. From this transformation we obtain the physical background
\bea\nonumber
\langle \Phi_{xy} \rangle & = & m( e^{f} E^+ + m\,x e^{-f} E^-) + \eps \left [\th_y\p_xf + \frac{m}{2}(mx {k}-{k}^*)\right ]P + \mu^2 (x - y) Q + \CO(\eps^2)\\
\langle A _{0,1}\rangle & = & -\frac{i}{2} \bar{\partial} f P - i \frac{\eps}{2}\left (\bar\p {k}\,e^{f}E^++\bar \p {k}^*\,e^{-f}E^-\right ) +  \CO(\eps^2)
\label{bkgrealeps}
\eea
Inserting (\ref{bkgrealeps}) into the D-term equation we recover that $f$ has again to satisfy the Painlev\'e equation (\ref{eq:comp}) while $k$ satisfies a more complicated differential equation given in Appendix \ref{ap:wave}. Using that near the origin $f = \log c+m^2c^2 x\bar{x}+\dots$ we find the solution
\be
k=\bar\th_y\, c^2mx+\th_y\,\frac{1-c^2}{2c^2-1}m^2 \bar x^2+\dots
\ee
where the dots stand for higher powers of $x,\bar{x}$. 

Finally, let us restore the presence of primitive fluxes (\ref{sumFp}). As these fluxes commute with all the other elements of the background their presence does not modify the discussion above, and we can add their contribution to the corrected background independently. At the ends one finds 
\bea
\label{bkgrealeps21}
\langle \Phi_{xy} \rangle & = & m( E^+ + m\,x E^-) + \eps \left [\th_y\p_xf + \frac{m}{2}(mx {k}-{k}^*)\right ]P \\ 
\nonumber
& & + \mu^2 (x - y) Q+\ \eps \left[\theta_y (\bar{x} Q_R - \bar{y} Q_S) + \theta_x (\bar{x} Q_S + \bar{y}Q_R)  \right]
 + \CO(\eps^2)\\
\langle A _{0,1}\rangle & = & \langle A_{0,1}^p \rangle -\frac{i}{2} \bar{\partial} f P - i \frac{\eps}{2}\left (\bar\p {k}\,e^{f}E^++\bar \p {k}^*\,e^{-f}E^-\right ) +  \CO(\eps^2)
\label{bkgrealeps22}
\eea
where $\langle A_{0,1}^p \rangle$ stands for the potential of the primitive flux (\ref{sumFp}) in a physical gauge. Notice that the $\CO(\eps)$ corrections to the worldvolume flux lie along the non-commuting generators $E^\pm$, while for the Higgs background they lie along the Cartan of $E_6$.

In the following we will solve for the wavefunctions that satisfy (\ref{eq:flucF}) and (\ref{eq:fluc2}) for the pair $(a,\vphi)$ and at first order in the non-perturbative parameter $\eps$. That is, we will be looking for solutions to the system
\begin{subequations}
\label{eq:flucFnp}
\begin{align}
\label{eq:fluc1np}
\pa a&=\CO(\eps^2)
\\\label{eq:fluc3np}
\pa\varphi-i[a,\lp]+\epsilon\partial\theta_0\,\w\,\partial_{\langle A\rangle}a&=\CO(\eps^2)
\\\label{eq:fluc2np}
\omega \wedge \ph a - \frac{1}{2}  [\langle \bar{\Phi} \rangle, \varphi]  & = \CO(\eps^2)
\end{align}
\end{subequations}
where $\langle A \rangle$ and $\langle \Phi \rangle$ are respectively specified by (\ref{bkgrealeps21}) and (\ref{bkgrealeps22}).

\subsubsection*{Sector 5}

The sector $\mathbf{5}$ is relatively simple due to the fact that its zero modes are not charged under the generators of the $\mathfrak{su(2)}$ algebra $\{E^\pm, P\}$. More precisely, for this sector $\langle A _{0,1}\rangle$ reduces to $\langle A _{0,1}^p\rangle$, and $\langle \Phi_{xy} \rangle$ to the second line of (\ref{bkgrealeps21}). As a result, solving the zero mode equations (\ref{eq:flucFnp}) for this sector is very similar to the analogous problem for the $SO(12)$ local model of \cite{fimr12}. Hence in the following we simply present the final result, and refer the reader to Appendix \ref{ap:wave} and section 5.1 of \cite{fimr12} for further details.

The solution to the non-perturbative zero mode equations is given by
\be
\label{wave5np}
\vec\vphi_{\mathbf{5}}\, =\, \gamma_{\mathbf{5}}\left ( \begin{array}{c}
i\frac{\zeta_{\mathbf{5}}}{2\mu^2}\\
i\frac{(\zeta_{\mathbf{5}}-\lam_{\mathbf{5}})}{2\mu^2}\\
1\end{array}\right )\, \chi_{\mathbf{5}}^{\rm np}, \
\qquad
\chi_{\mathbf{5}}^{\rm np} = 
e^{\frac{q_R}{2}(|x|^2-|y|^2)-q_S (x \bar y +y\bar x)+(x-y)(\zeta_{\mathbf{5}}\bar x-(\lam_{\mathbf{5}}-\zeta_{\mathbf{5}})\bar y))}(1+\eps\Upsilon_{\mathbf{5}})
\ee
with $\lam_{\mathbf{5}}$, $\zeta_{\mathbf{5}}$ defined as in (\ref{wave5p}). The $\CO(\eps)$ non-perturbative correction is
\be
\Upsilon_{\mathbf{5}}=-\frac{1}{4\mu^2}(\zeta_{\mathbf{5}} \bar{x} - (\lam_{\mathbf{5}} - \zeta_{\mathbf{5}})\bar{y})^2(\th_x+\th_y)+\frac{\delta_1}{2}(x-y)^2+\frac{\delta_2}{\zeta_{\mathbf{5}}}(x-y)(\zeta_{\mathbf{5}} y+(\lam_{\mathbf{5}}-\zeta_{\mathbf{5}})x)
\ee
with the constants $\delta_1$, $\delta_2$ given by (\ref{d1}) and (\ref{d2}) respectively.
As in \cite{fimr12} one can check that the corrections to the norm (\ref{norm5}) only appear at $\CO(\eps^2)$, because $\CO(\eps)$ terms that appear in the integrand of (\ref{4dnorm5}) are not invariant under the rotation $(x,y) \raw e^{i\a}(x,y)$.

\subsubsection*{Sector 10}

Similarly to the case of perturbative zero modes, finding the non-perturbative corrections to the wavefunctions of the sector $\mathbf{10}$ is in general rather involved. Nevertheless, taking the same approximations as in the perturbative case, one may understand how this corrections look like and argue that they will not be relevant for computing the matrix of physical Yukawa couplings. 

The first step is to switch off the primitive fluxes and realize that, in the same way that $a =\bar{\p}\xi$ and (\ref{genphi3}) solve the F-term equations (\ref{eq:fluc1np}) and (\ref{eq:fluc3np}) in the holomorphic gauge, in the real gauge they are satisfied by
\begin{subequations}
\label{Fsolnp}
\begin{align}
\label{asolnp}
a & =  g\, \bar\p \xi\\
\label{phisolnp}
\vphi & =  g \left(h - i \Psi \xi - \eps \p \th_0  \wedge \p \xi \right) \, =\, g\, U\, dx \wedge dy
\end{align}
\end{subequations}
with $g$ given by (\ref{cgtnp}) and $\Psi$ given by (\ref{cocacola}). Here $a$, $\vphi$, $\chi$ are $SU(2)$ doublets as in eq.(\ref{Fsol}). The same applies to $U$, which can be expanded in powers of $\eps$ as
\be
U\, =\, U^{(0)} + \eps\, U^{(1)} + \, \CO(\eps^2)
\ee
where $U^{(0)}$ corresponds to solution found for $\eps = 0$, namely
\be
 U^{(0)}_- \, =\,  e^{\lam_{\mathbf{10}}x\bar{x}} h (y) \qquad \quad  U^{(0)}_+ \, =\, 0
\ee
Then, similarly to (\ref{solxi}) one may solve for $\xi$ as 
\be
\begin{array}{c}
\xi\, =\, \xi^{(0)} +  i \eps \Psi^{-1} \left[U^{(1)} + \p_x\th_0 \p_y\xi^{(0)} - \p_y\th_0\p_x\xi^{(0)} \right] + \CO(\eps^2) \\
 \xi^{(0)} \, =\, i \Psi^{-1} (U^{(0)} - h)
\end{array}
\label{solxinp}
\ee
and then solve for $U^{(1)}$ by inserting this expression into the D-term equation (\ref{eq:fluc2np}). As in the perturbative case this problem can be easily solved in the limit $\mu \raw 0$,  obtaining that $U^{(1)}_- =0$. As a result, in this limit we have the structure
\be
\xi_+ \, =\, \xi_+^{(0)} + 0 + \CO(\eps^2) \qquad \quad \xi_- \, =\, 0 + \eps\, \xi_-^{(1)} + \CO(\eps^2)
\ee
that is, the $\CO(\eps)$ corrections to $\xi$ are contained in the opposite doublet as the tree-level contribution. The same statement applies to $a$ and $\vphi$. Indeed, we have that
\be
\vphi_{xy}\,=\, g^{(0)} U^{(0)} + \eps (g^{(0)} U^{(1)} + g^{(1)} U^{(0)}) + \CO(\eps^2)
\label{phi10np}
\ee
where we have decomposed $g = g^{(0)} + \eps g^{(1)} + \CO(\eps^2)$ as in (\ref{cgtnp}). Then, because $g^{(0)}$ only involves $P$ and $g^{(1)}$ involves $E^\pm$ we have 
\be
\vphi_+ \, =\, 0 + \eps\, \vphi_+^{(1)} + \CO(\eps^2) \qquad \quad \vphi_- \, =\, \vphi_-^{(0)} + 0 + \CO(\eps^2)  
\ee
Finally, a similar argument shows that $a_+ = a_+^{(0)} + \CO(\eps^2)$ and $a_- = a_-^{(1)} + \CO(\eps^2)$ and so the wavefunctions (\ref{phys10}) have a correction of the form
\be
\label{phys10np} 
\vec{\vphi}_{\mathbf{10}^+} \, =\, 
\left(\begin{array}{c} \bullet \\ \bullet \\ 0\end{array}\right)
 + \eps
\left(\begin{array}{c} 0 \\ 0 \\ \bullet\end{array}\right) +\CO(\eps^2)
 \qquad 
\vec{\vphi}_{\mathbf{10}^-} \, =\,
\left(\begin{array}{c} 0 \\ 0 \\ \bullet\end{array}\right)
 + \eps
 \left(\begin{array}{c} \bullet \\ \bullet \\ 0\end{array}\right)+\CO(\eps^2)
\ee
One can check that this structure remains even after we restore the presence of non-primitive fluxes. Then, since the $\CO(\eps)$ correction vector is orthogonal to the 0$^{\rm th}$-order solution, it is easy to see that no $\CO(\eps)$ correction to the normalization factors $\g_{\mathbf{10}}^j$ arises by plugging these corrected wavefunctions into (\ref{4dnorm10}).

\section{Physical Yukawas and mass hierarchies}
\label{s:physyuk}

Given the above solutions for the non-perturbative wavefunctions one can insert them into (\ref{yukubic}) and compute their triple overlap to obtain the matrix of physical Yukawa couplings, that is the Yukawas in a basis where 4d kinetic terms are canonically normalized. 

As we will see below the final result for the U-quark Yukawa matrix is
\be
Y_U\, =\, \frac{\pi^2\g_{\mathbf{5}}}{4\rho_\mu\rho_m}
\left(
\begin{array}{ccc}0&0& \tilde \eps \rho_\mu^{-1} \g_{L}^1\g_{R}^3 \\0& \ \tilde \eps \rho_\mu^{-1} \g_{L}^2\g_{R}^2 &0\\ \tilde \eps \rho_\mu^{-1} \g_{L}^3\g_{R}^1 &0&- 2 \gamma_{L}^3 \g_{R}^3
\end{array}
\right)
+\mathcal{O}(\tilde\eps^2)
\label{Yukphys}
\ee
where
\be
\rho_{\mu} \, =\, \frac{\mu^2}{m_*^2} \qquad \qquad \rho_m  \, =\, \frac{m^2}{m_*^2} \qquad \qquad  \tilde \eps\, =\, \eps\, (\th_x  + \th_y)
\label{paramyp}
\ee
are all flux-independent parameters. The worldvolume flux dependence (and in particular the hypercharge dependence) is encoded in the normalization factors $\g_{\mathbf{5}}$ and $\g_{R,L}^i$, where $\g_{\mathbf{5}}$ is given by (\ref{norm5}) with the values of $q_R$, $q_S$ for the sector $\mathbf{5}_1$ of table \ref{t:sectors}. Finally, $\g_{R}^i$ is given by (\ref{norm10}) using the values of $q_R$ and $q_S$ in the first row of table \ref{t:sectors}, and similarly for $\g_L^i$ with the values in the second row. 

We would like to see if this structure for Yukawa couplings allows to fit experimental fermion masses. Since our expressions apply at the GUT scale, presumably of order $10^{16}$ GeV, the data need to be run up to this scale. Table \ref{t:masses} shows the result of doing so for the MSSM quark mass ratios, for different values of $\tan\beta$ as taken from ref.\cite{Ross:2007az}. In the following we will analyze if this  spectrum can be accommodated in our scheme. 

\begin{table}[htb] 
\renewcommand{\arraystretch}{1.25}
\begin{center}
\begin{tabular}{|c||c|c|c|}
\hline
tan$\beta$  &  10&   38  &  50 \\
\hline\hline
$m_u/m_c$ &   $2.7\pm 0.6\times 10^{-3}$   &  $2.7\pm 0.6\times 10^{-3}$&$2.7\pm 0.6\times 10^{-3}$  \\
\hline
$m_c/m_t$ &   $2.5\pm 0.2\times 10^{-3}$ &$2.4\pm 0.2\times 10^{-3}$&$2.3\pm 0.2\times 10^{-3}$ \\
\hline\hline
$m_d/m_s$ &   $5.1\pm 0.7\times 10^{-2}$   &  $5.1\pm 0.7\times 10^{-2}$  & $5.1\pm 0.7\times 10^{-2}$  \\
\hline
$m_s/m_b$ &    $1.9\pm 0.2\times 10^{-2}$   &  $1.7\pm 0.2\times 10^{-2}$  & $1.6\pm 0.2\times 10^{-2}$  \\
\hline\hline
$Y_t $  &    $0.48\pm0.02 $   &  $0.49\pm0.02 $ &    $0.51\pm0.04 $ \\
\hline
$Y_b $  &    $0.051\pm0.002 $   &  $0.23\pm0.01 $ &    $0.37\pm0.02 $ \\
\hline
\end{tabular}
\end{center}
\caption{\small Running mass ratios of quarks at the unification scale and for different values of $tan$ $\beta$, as taken from ref.\cite{Ross:2007az}.
The Yukawa couplings $Y_{t,b}$ at the unification scale are also shown.}
\label{t:masses}
\end{table}

\subsection{The physical Yukawa matrix}

Let us first perform the computation of the physical Yukawa matrix. Inserting the zero-mode wavefunctions for the $\mathbf{5}$ and $\mathbf{10}$ sector into the cubic coupling (\ref{yukubic}) and applying the $E_6$ group theory relations we obtain
\be
Y_U^{ij}\, =\,
m_*^4 \int_S 
 \mathrm{det}\left(\vec\vphi_{\mathbf{5}}, \vec{\vphi}_{\mathbf{10}^M}^i , \vec{\vphi}_{\mathbf{10}^N}^j  \right) \eps_{MN}\, {\rm dvol}_S
 \label{yukdet}
\ee
with $M, N = \pm$ and $\eps_{MN}$ the $\mathfrak{su(2)}$ antisymmetric tensor. To obtain the Yukawas at zeroth order in $\epsilon$ we just need to plug into (\ref{yukdet}) the perturbative wavefunctions computed in subsection \ref{ss:zmp}. One then finds the expression
\be
Y_U^{(0)}{}^{ij}\, =\, 2  m_*^4 \, \g_{\mathbf{5}} \g_{\mathbf{10}}^i \g_{\mathbf{10}}^j \,
\mathrm{det}\left(\vec v_{\mathbf{5}}, \vec{v}_{\mathbf{10}^+}^i , \vec{v}_{\mathbf{10}^-}^j  \right)
\int_S \chi_{\mathbf{5}} \chi_{\mathbf{10}}^i  \chi_{\mathbf{10}}^j\, {\rm d vol}_S
\ee
where the vectors $\vec{v}_\rho$ are defined as in (\ref{vec5}) and (\ref{vec10}), and $\chi_\rho$ are the perturbative scalar wavefunctions of (\ref{wave5p}) and (\ref{wave10p}). As the product of these three wavefunctions is sharply localized around the origin one can replace the domain of integration by $\IC^2$. Taking the holomorphic representatives for each family as in (\ref{holorep}) one obtains
\be
Y_U^{(0)}{}^{33}\, =\, - \frac{\pi^2}{2\rho_\mu\rho_m} \g_{{\mathbf{5}}}\gamma_{L}^3 \g_{R}^3 \qquad \qquad Y_U^{(0)}{}^{ij}\, =\, 0 \quad {\rm for} \quad i\neq 3 \neq j
\label{topyuk}
\ee
where the slope densities $\rho_{m,\mu}$ are defined as in (\ref{paramyp}), $\g_{\mathbf{5}}$ is the normalization factor (\ref{norm5}) evaluated for the sector $\mathbf{5}_1$ of table \ref{t:sectors} and $\g_{R,L}^i$ are the normalization factors (\ref{norm10}) evaluated for the sectors $\mathbf{10}_{1,2}$ of the same table. Then, as expected from our construction, obtains a rank 1 Yukawa matrix at the perturbative level, which moreover is in perfect agreement with the result of the residue computation of Section \ref{s:holoyuk}.

The $\CO(\eps)$ contribution to (\ref{yukdet}) can be written as
\bea\nonumber
Y_U^{(1)}{}^{ij} & = &
2 m_*^4 \int_S 
\left[ \mathrm{det}(\vec\vphi_{\mathbf{5}}^{(1)}, \vec{\vphi}_{\mathbf{10}^+}^{(0)\, i} , \vec{\vphi}_{\mathbf{10}^-}^{(0)\, j}) 
 +  \mathrm{det}(\vec\vphi_{\mathbf{5}}^{(0)}, \vec{\vphi}_{\mathbf{10}^+}^{(1)\, i} , \vec{\vphi}_{\mathbf{10}^-}^{(0)\, j})\right. \\
 & & \qquad \qquad \left. +\,  \mathrm{det}(\vec\vphi_{\mathbf{5}}^{(0)}, \vec{\vphi}_{\mathbf{10}^+}^{(0)\, i} , \vec{\vphi}_{\mathbf{10}^-}^{(1)\, j})\right]  
{\rm d vol}_S
\label{yukdet1}
\eea
where we have split the corrected wavefunction (\ref{wave5np}) as $\vec\vphi_{\mathbf{5}} = \vec\vphi_{\mathbf{5}}^{(0)} + \eps\, \vec\vphi_{\mathbf{5}}^{(1)}$ and similarly for the wavefunctions $\vec{\vphi}_{\mathbf{10}^\pm}^{i} = \vec{\vphi}_{\mathbf{10}^\pm}^{(0)\, i} + \eps \vec{\vphi}_{\mathbf{10}^\pm}^{(1)\, i} $ in the sector $\mathbf{10}$. 

Performing the integral of the first term in (\ref{yukdet1}) one obtains
\be
Y_U^{(1)}\, =\, \frac{\pi^2\g_{\mathbf{5}}}{4\rho_\mu^2\rho_m}
\left(
\begin{array}{ccc}0&0& \g_{L}^1\g_{R}^3 \\0&   \g_{L}^2\g_{R}^2 &0\\ \g_{L}^3\g_{R}^1 &0 & 0 
\end{array}
\right)\, (\th_x+ \th_y) 
\label{yukphys1}
\ee
matching the result (\ref{Yukhol}) for the case (\ref{choiceab}) that we are considering. Recall that in the computations of Section \ref{s:holoyuk} the $\CO(\eps)$ corrections to the Yukawa matrix came entirely from the corrections to the wavefunction in the $\mathbf{5}$ sector, while the corrections to the $\mathbf{10}$ sector did not contribute to the Yukawas. One can argue that the same will happen here as follows. First notice that due to the zero mode structure (\ref{phys10np}), $\vec{\vphi}_{\mathbf{10}^-}^{(0)\, j}$ and $\vec{\vphi}_{\mathbf{10}^+}^{(1)\, i}$ are proportional to each other up to multiplication by a complex function, and so the determinant in the second term of (\ref{yukdet1}) vanishes  identically. Second, for $\zeta_{\mathbf{10}} = 0$ the tree-level wavefunctions $\vec{\vphi}_{\mathbf{10}^+}^{(0)\, i}$ only have one non-vanishing component (c.f.(\ref{phys10})), and one can then see that the same applies to $\vec{\vphi}_{\mathbf{10}^-}^{(1)\, i}$, so that the third determinant in (\ref{yukdet1}) vanishes as well. For $\zeta_{\mathbf{10}} \neq 0$ such determinant may not vanish identically, but its integral should vanish because $\zeta_{\mathbf{10}}$ is proportional to the flux $q_S$ and the integral (\ref{yukdet}) should not depend explicitly on background worldvolume fluxes. Indeed, recall that (\ref{yukdet}) is equivalent to (\ref{yukubic}), which is invariant under complexified gauge transformations. Such transformations can be used to gauge away any dependence on the worldvolume flux, and so the result obtained for $q_S=0$ should be true in general. In fact, one can use a complexified gauge transformation to take the wavefunctions computed in the previous section to the ones used in Section \ref{s:holoyuk} in the residue formula, which is why both results match.\footnote{In relating wavefunctions by a complexified gauge transformation we assume that the definition of families in terms of  monomials is preserved. That is, we assume that the choice of family representative (\ref{holorep}) is mapped to (\ref{solh10}) by a complexified gauge transformation that removes the flux dependence from the wavefunction.} 

Finally, adding up these two results as
\be
Y_U\, =\, Y_U^{(0)} + \eps\, Y_U^{(1)} + \CO(\eps^2)
\ee
we obtain (\ref{Yukphys}), as claimed above. In the following we will analyze if given these up-like Yukawas we can reproduce the data in table \ref{t:masses}.

\subsection{The top quark Yukawa}

The Yukawa for the top quark is given by the 33 entry of (\ref{Yukphys}). To analyze its value it is useful to express the quantities $\rho_\mu$ and $\rho_m$ as
\be
\rho_\mu \, =\, \left( \frac{\mu}{m_*} \right)^2\, =\, (2\pi)^{3/2} g_s^{1/2} \sigma_\mu \qquad \qquad 
\rho_m \, =\, \left( \frac{m}{m_*} \right)^2\, =\, (2\pi)^{3/2} g_s^{1/2} \sigma_m
\ee
where $\sigma_{\mu} = (\mu/m_{st})^2$ and $\sigma_{m} = (m/m_{st})^2$ are the 7-brane intersection slopes measured in units of $m_{st}$, the scale that in the type IIB limit reduces to the string scale $m_{st} = 2\pi \a'$ and which is related to the F-theory scale as $m_{st}^4 = g_s(2\pi)^3m_*^4$ \cite{fimr12}. We then have that
\be
|Y_t|\, =\,  (8\pi g_s)^{1/2} \sig_m c\
\tilde{\g}_{{\mathbf{5}}_1} \tilde{\g}_{\mathbf{10}_1} \tilde{\g}_{\mathbf{10}_2}
\label{topyuk}
\ee
where
\bea
\tilde{\g}_{{\mathbf{5}}_1} & = & \left(- \frac{(2\zeta_{{\mathbf{5}}_1}+q_R^{{\mathbf{5}}_1})(q_R^{{\mathbf{5}}_1}+2\zeta_{{\mathbf{5}}_1}-2\lam_{{\mathbf{5}}_1})+(q_S^{{\mathbf{5}}_1}+\lam_{{\mathbf{5}}_1})^2}{4\mu^4+\zeta_{{\mathbf{5}}_1}^2+(\zeta_{{\mathbf{5}}_1}-\lam_{{\mathbf{5}}_1})^2}\right)^{1/2}\\
\tilde{\g}_{\mathbf{10}_i} & = & \left( -\frac{q_R^{\mathbf{10}_i}}{\frac{m^4}{2\lam_{\mathbf{10}_{i}}+q_R^{\mathbf{10}_{i}} (1+\zeta_{\mathbf{10}_{i}}^2)-m^2 c^2}+\frac{c^2\lam_{\mathbf{10}_{i}}^2}{2\lam_{\mathbf{10}_{i}}+q_R^{\mathbf{10}_{i}} (1+\zeta_{\mathbf{10}_{i}}^2)+m^2 c^2}}\right)^{1/2} \quad i=1,2
\eea
with $q_{R,s}^{\mathbf{10}_i}$, $i=1,2$ the values of $q_{R,S}$ in the $i^{\rm th}$ row of table \ref{t:sectors}, $q_{R,S}^{{\mathbf{5}}_1}$ the ones in the fourth row, etc.
Notice that this expression is quite similar to the one obtained for the third generation of down-like Yukawas in \cite{fimr12} (c.f. eq.(7.8) there) , except for an extra factor of $\sqrt{2}c$ which for the value (\ref{nopolec}) is very close to 1. Hence in principle one expects that the Yukawa of the top and of the bottom are of the same order of magnitude, which in the scheme of the MSSM would favor a large tan$\beta$. 

From (\ref{topyuk}) one may proceed as in \cite{fimr12} and estimate that primitive worldvolume flux densities are of the order
\be
M, N \, \simeq\, 0.29\, g_s^{1/2} m_{st}^2 
\ee
with $g_s$ not too small. In fact, the diluted flux approximation is one of the requirements that we need to impose in order to be able to trust the 7-brane effective action that led to the zero mode equations of Section \ref{s:zeromodes}. A further self-consistency restriction comes from the fact that the non-primitive flux (\ref{sumFnp}) must be slowly varying in the region where wavefunctions are localized. As discussed in Appendix \ref{ap:wave}, this leads to the condition (\ref{condP2}). Finally, recall that in order to simplify the $\mathbf{10}$-plet zero mode equations restricted ourselves to the region of the parameter space such that $m \gg \mu$. All these approximations are only important for computing the normalization factors for the wavefunctions, while the computation of holomorphic Yukawas in Section \ref{s:holoyuk} is independent of them.

Given these restrictions one can see that one may accommodate a realistic value for the Yukawa of the top at the unifications scale. Indeed, if one for instance takes the values (in units of $m_{st}$)
\be
M= 0.3\,, \quad N= 0.03\,, \quad \tilde N_Y = 0.6 \,, \quad N_Y=-0.18 \,, \quad m= 0.5 \,, \quad \mu = 0.1\,, 
\label{paramval}
\ee
with  $g_s = 1$ and $c$ as in (\ref{nopolec}) one obtains 
\be
Y_t= 0.5
\ee
in quite good agreement with the values of table \ref{t:masses}. One can also check that the wavefunctions are sufficiently localized in a region where the first two terms of (\ref{Painori}) are a good approximation for the Painlev\'e transcendent.

\subsection{Up-type quarks mass hierarchies}

In order to analyze the flavor  hierarchies among different U-quarks let us consider the matrix
\be
\frac{Y_U}{Y^{33}}\, =\, 
\left(
\begin{array}{ccc}0&0& -\oh \tilde \eps \rho_\mu^{-1} \frac{\g_{L}^1}{\g_L^3} \\0& -\oh \tilde \eps \rho_\mu^{-1} \frac{\g_{L}^2\g_{R}^2}{\g_L^3\g_R^3} &0\\ - \oh \tilde \eps \rho_\mu^{-1} \frac{\g_{R}^1}{\g_L^3} &0&1
\end{array}
\right)
+\mathcal{O}(\tilde\eps^2)
\label{Yukrat}
\ee
whose eigenvalues are
\bea\nonumber
\lam_1&=&1+\CO(\eps^2)\\\nonumber
\lam_2&=& - \eps \frac{1}{2\rho_\mu}\frac{\g_{L}^2\g_{R}^2}{\g_L^3\g_R^3}  ( \theta_{x}+{\theta}_{y})+\CO(\eps^2)\\\nonumber
\lam_3&=&\CO(\eps^2).
\eea
where we have used the expression for $\tilde\eps$ in (\ref{paramyp}). This yields automatically a hierarchy of U-quark masses of the form $(1, \eps, \eps^2)$ in fact quite similar to the one found in \cite{fimr12} for the D-quarks and leptons. As in there, the quotient of quark masses of different families is rather simple. Namely identifying the first and second eigenvalues with the third and second generations of U-quarks we have 
\bea
\frac{m_c}{m_t} & = & \oh \left (\frac{q_R^{\mathbf{10}_1}q_R^{\mathbf{10}_2}}{\mu^4} \right )^{1/2}\eps\, (\th_{x}+ \th_{y})\\ \nonumber
&  = & \oh \frac{M}{\mu^2} \left(1 + \frac{2\tilde{N}_Y}{3M}\right)^{1/2} \left(1 - \frac{\tilde{N}_Y}{6M}\right)^{1/2} \eps\, (\th_{x}+ \th_{y})
\eea
where we have used that $q_R^{\mathbf{10}_1} = M+\frac{2}{3}\tilde N_Y$ and $q_R^{\mathbf{10}_2} =M-\frac{1}{6}\tilde N_Y$. Hence it is quite easy to accommodate the hierarchy between the charm and the top quark with a small non-perturbative parameter $\eps$. In fact, one may consider the ratio of flux densities $\tilde N_Y/ M \sim 1.8$ obtained in \cite{fimr12} for the down-like Yukawa point $p_{\rm down}$ and apply it to this expression, since, if the two Yukawa points $p_{\rm up}$ and $p_{\rm down}$ are not far away the flux densities should be alike. One then obtains that a realistic mass ratio requires
\be
\frac{M}{\mu^2} \eps\, (\th_{x}+ \th_{y}) \simeq 4 \times 10^{-3}
\ee
which can be achieved by taking $\tilde \eps = \eps\, (\th_{x}+ \th_{y}) \sim 10^{-4}$ as in \cite{fimr12} and $M$ and $\mu$ as in (\ref{paramval}). Of course a more detailed analysis would require to embed both Yukawa points $p_{\rm up}$ and $p_{\rm down}$ in the same local model, possibly in a region of $E_7$ or $E_8$ enhancement. We leave such analysis for future work. 

\section{Conclusions and outlook}
\label{s:conclu}

In this paper we have analyzed the structure of up-like Yukawa couplings in F-theory models of SU(5) unification, taking into account the contribution of non-perturbative effects. More precisely, we have considered an explicit local model based on a T-brane background and such that up-like $\mathbf{10\times 10 \times 5}$ Yukawa couplings are generated at the intersection of a $\mathbf{10}$ and a $\mathbf{5}$ matter curve. From the general results of \cite{cchv10} one expects to obtain a rank one matrix of Yukawas from such configuration, as we have verified in our model. We have then incorporated an extra ingredient to this local model, namely the presence of non-perturbative effects sourced by distant 4-cycles of the compactification, along the lines of \cite{mm09}. As shown in \cite{afim,fimr12} one may easily incorporate these effects into the ultra-local approach that allows to compute Yukawa couplings, with the general result that the rank of the Yukawa matrix is enhanced from one to three. 

We have seen that this statement remains true for up-like Yukawas, by computing the non-perturbative corrections to these couplings in a local $E_6$ model based on T-branes. Moreover, we have obtained an U-quark mass hierarchy of the form $(\CO(\eps^2), \CO(\eps), \CO(1))$, as an explicit computation of the Yukawa matrix shows. This hierarchy is already manifest at the level of the holomorphic Yukawa couplings, which we have computed via a generalized residue formula, and is also recovered in the matrix of physical Yukawas computed via wavefunction overlap. In fact, this hierarchical structure is similar to the one obtained in \cite{fimr12} in the context of $\mathbf{10 \times \bar{5} \times \bar{5}}$ Yukawa couplings, and  it allows to reproduce ratios of quark and lepton masses compatible with experiment once that the flux dependent normalization factors of the different wavefunctions are taken into account. Finally, we have verified that in this scheme the physical Yukawa for the top is of the right order of magnitude, which is the main motivation to consider SU(5) F-theory GUT's as opposed to their type II analogues. 

Computing zero mode wavefunctions for a T-brane background is in general a rather involved task, mainly because the background itself is described by complicated non-linear differential equations like Painlev\'e equations. These complications however disappear when computing holomorphic Yukawas, as our results of Section \ref{s:holoyuk} show. In this respect, one can conclude that the hierarchical structure obtained in eq.(\ref{Yukhol}) is a rather robust prediction of this class of SU(5) F-theory models. An intriguing result that we have obtained in this context is that the non-perturbative contributions to the Yukawa couplings arise from the $\CO(\eps)$ corrections to the internal wavefunction for the up-Higgs $H_u$. In fact, a rank 3 matrix of Yukawas is generated only if the holomorphic function $\theta_0$ that locally describes the non-perturbative effects is non-constant along the Higgs curve $\mathbf{5}_H$. It would be interesting to have a deeper understanding of such result, and what does it implies for a global completion of our local model. 

The full complexity of the T-brane background becomes important when computing zero mode wavefunctions in the $\mathbf{10}$ matter curve, and in particular for the $\CO(\eps)$ corrections to these wavefunctions. We have however found that applying the limit $\mu/m \raw 0$ to the T-brane background (\ref{sumPhi}) drastically simplifies these zero mode equations and allows to solve for them analytically. It would be very interesting to extend our results away from this limit, although the fact that we recover the same Yukawa matrix as in the residue computation hints that our results capture most of the relevant physics. It would then be interesting to see if these wavefunctions near the $E_6$ point can be used to compute other quantities of phenomenological interest in F-theory GUT's, like higher dimensional operators that trigger proton decay \cite{cdp11,nosusy} and soft SUSY-breaking terms \cite{civ13}.

Besides computing physical Yukawas for arbitrary $\mu$ our results can be extended in a number of ways.  First one could consider the contribution of further terms in the expansion (\ref{suponpintro}) that describe the non-perturbative effects. From the results of \cite{fimr12} terms proportional to $\theta_n$ with $n>0$ should not alter the hierarchical structure that we have obtained here, but they could be important in order to compute explicitly the Yukawa couplings of the lightest generation. Such task would in addition involve extending our computations to $\CO(\eps^2)$ in the non-perturbative parameter $\eps$, which would also be important to compute the CKM matrix for these models. In fact, because the CKM matrix involves considering both Yukawa points $p_{\rm up}$ and $p_{\rm down}$ simultaneously, it would make sense to consider local models of $E_7$ or $E_8$ enhancement to implement our approach \cite{cftw11}. Finally, it would be interesting to see how the ultra-local parameters that determine the Yukawa couplings are realized in local and global completions of our F-theory SU(5) model \cite{collinucci09,bgjw09,mss2,Cordova,mss,gkw09,Knapp:2011wk}. We hope to return to these points in the future.

\bigskip

\bigskip

\centerline{\bf \large Acknowledgments}

\bigskip

We thank H.~Hayashi, L.~E.~Ib\'a\~nez, E.~Palti, S.~Sch\"afer-Nameki and  S. Theisen for useful discussions. 
A.F. is grateful to the AEI-Potsdam, the CERN TH Division, and the IFT-UAM/CSIC 
for hospitality and support at various stages of this work.
F.M., D.R. and G.Z. thank HKUST IAS for hospitality and support during progress of this work.
This work has been partially supported by the grants FPA2009-07908 and FPA2012-32828 from the MINECO, HEPHACOS-S2009/ESP1473 from the C.A. de Madrid, the REA grant agreement PCIG10-GA-2011-304023 from the People Programme of FP7 (Marie Curie Action), the ERC Advanced Grant SPLE under contract ERC-2012-ADG-20120216-320421 and the grant SEV-2012-0249 of the ``Centro de Excelencia Severo Ochoa" Programme. F.M. is supported by the Ram\'on y Cajal programme through the grant RYC-2009-05096. D.R. is supported through the FPU grant AP2010-5687. G.Z. is supported through a grant from ``Campus Excelencia Internacional UAM+CSIC".


\appendix



\section{Wavefunctions at the $E_6$ Yukawa point}
\label{ap:wave}

In this appendix we will present the strategy pursued to solve the equations of motion for the fluctuations in a real gauge that allow to compute the normalization factors. We will see that the computation for the $\mathbf{10}$ sector is quite involved due to the T-brane background and in order to find an analytic solution we restrict to some particular region of parameters. In contrast the solution for the $\mathbf{5}$ sector can be computed exactly since it does not feel the non-Abelian part of the Higgs background.\footnote{The computation for the $X,Y$-boson wavefunctions at the bottom of table \ref{t:sectors} proceeds exactly as in Appendix A of \cite{fimr12} and so will not be repeated here.}

\subsection*{Zero modes for the $\mathbf{5}$ sector}

For the $\mathbf 5$ sector we follow closely the computations in \cite{afim} and \cite{fimr12}. We recall from the main text that the equations of motion for the fluctuations can be written as
\be\label{Dir:mas}
\mathbf{D}_A \Psi=0\,,
\ee
which are reminiscent of a Dirac equation. 
To solve (\ref{Dir:mas}) it is convenient to take its modulo square for it is possible to decompose the operator $\mathbf{D}^\dagger_A \mathbf{D}_A$ as
\be
\mathbf{D}^\dagger_A \mathbf{D}_A = - \Delta \mathbf{1}_4+\mathbf{M}\,,
\ee 
where the Laplacian $\Delta$ is defined as $\Delta = \{D_x,D_{\bar x} \}+\{D_y,D_{\bar y} \}+\{D_z,D_{\bar z} \}$ and the matrix $\mathbf{M}$ will depend on the worldvolume fluxes and intersection slopes.
Whenever the flux matrix $\mathbf{M}$ and the Laplacian commute (for instance this happens in the case of constant fluxes and abelian Higgs) it is possible to diagonalize simultaneously these two operators. We will start by diagonalizing 
the operator $\mathbf{M}$ and then use its eigenmodes to solve the complete set of equations. As an aside we mention that the strategy outlined so far is general and not restricted to the $\mathbf{5}$ sector. The issue in the
$\mathbf{10}$ sector that spoils its efficiency is that the operator $\mathbf{M}$ will be function of non-constant fluxes and thus it will not commute with the Laplacian.\\
For the $\mathbf{5}$ sector the operator $\mathbf{M}$ has the form

\be\label{flux5}
\mathbf M_{\mathbf{5}}=\left ( \begin{array}{cccc}
0&0&0&0\\0&-q^{\mathbf{5}}_R&q_S^{\mathbf{5}}&2i\mu^2\\
0&q_S^{\mathbf{5}}&q_R^{\mathbf{5}}&-2i\mu^2\\
0&-2i\mu^2&2i\mu^2&0\end{array}\right )= \left(\begin{array}{c c}0&0\\0& \mathbf{m}_{\mathbf{5}}\end{array}\right)\,.
\ee

The eigenvalues of the matrix $\mathbf{m}_{\mathbf{5}}$ are solutions of the secular equation
\be\label{sec:lam}
\lam^3_{\mathbf{5}}-(8\mu^4+(q_R^{\mathbf{5}})^2+(q_S^{\mathbf{5}})^2)\lam_{\mathbf{5}}+8\mu^4q_S^{\mathbf{5}}=0\,.
\ee
We will call the solutions of (\ref{sec:lam}) $\lam_{\mathbf{5}}^i$  chosen to satisfy the inequality $\lam^1_{\mathbf{5}} < \lam^2_{\mathbf{5}} < \lam^3_{\mathbf{5}}$. The eigenvectors of $\mathbf{m}_{\mathbf{5}}$ are
\be
\vec v_i=\left ( \begin{array}{c}
i\frac{\zeta^i_{\mathbf{5}}}{2\mu^2}\\
i\frac{(\zeta^i_{\mathbf{5}}-\lam^i_{\mathbf{5}})}{2\mu^2}\\
1\end{array}\right )\,, \quad \zeta^i_{\mathbf{5}} = \frac{\lam^i_{\mathbf{5}}(\lam^i_{\mathbf{5}}-q_R^{\mathbf{5}}-q_S^{\mathbf{5}})}{2(\lam^i_{\mathbf{5}}-q_S^{\mathbf{5}})}\,.
\ee
Knowing the form of the eigenvectors of $\mathbf{M}_{\mathbf{5}}$ we can look for a solution of (\ref{Dir:mas}) of the form
\be
\Psi_{\mathbf{5}} = \left(\begin{array}{c} 0 \\ \vec v_{\mathbf{5}} \end{array}\right)\chi_{\mathbf{5}}E_{\mathbf{5}}\,.
\label{vec5}
\ee
with $\vec v_{\mathbf{5}} = \vec v_1$. Plugging this in (\ref{Dir:mas}) we have that the function $\chi_{\mathbf{5}}$ has to satisfy
\begin{subequations}
\begin{align}
\left[-\zeta^1_{\mathbf{5}} D_x+(\lam_{\mathbf{5}}^1-\zeta_{\mathbf{5}}^1)D_y +2i \mu^2 D_z\right] \chi_{\mathbf{5}} & =0\,,\\
\left[2i \mu^2 D_{\bar y}+(\zeta_{\mathbf{5}}^1-\lam_{\mathbf{5}}^1)D_{\bar z}  \right]\chi_{\mathbf{5}} & =0\,,\\
\left[2i \mu^2 D_{\bar x}+\zeta_{\mathbf{5}}^1D_{\bar z}  \right]\chi_{\mathbf{5}} & =0\,.
\end{align}
\end{subequations}
The solution of this system of equations is
\be
\chi_{\mathbf{5}} = e^{\frac{q_R}{2}(|x|^2-|y|^2)-q_S (x \bar y +y\bar x)+(x-y)(\zeta_{\mathbf{5}}\bar x-(\lam_{\mathbf{5}}-\zeta_{\mathbf{5}})\bar y))} g_{\mathbf{5}}\left((\lam_{\mathbf{5}}^1-\zeta_{\mathbf{5}}^1)x+\zeta_{\mathbf{5}}^1y\right)\,,
\ee
where $g_{\mathbf{5}}$ is a holomorphic function.


\subsection*{Zero modes for the $\mathbf{10}$ sector}
We recall here the main strategy outlined in the main text that will allow us to find a solution to the intricate equations for the fluctuations in the $\mathbf{10}$ sector. We start finding the general solution to the F-term equations
that take the following form
\begin{subequations}\label{fluct}
\begin{align}\label{a1}
a &= e^{f P/2} \bar \p \xi\,,\\\label{a2}
\varphi &= e^{f P/2}( h - i \Psi \xi)\,,
\end{align}
\end{subequations}
where $h$ is a doublet of holomorphic functions and $\xi$ is a doublet of regular functions. Note that the fluctuations in (\ref{fluct}) will solve the F-term equations in holomorphic gauge for the primitive fluxes:
once we know the fluctuations it will be easy to pass to a real gauge
\be
a^{real} = e^{\frac{q_R}{2}(|x|^2-|y|^2)-q_S (x \bar y +y\bar x)} \,a\,, \quad \varphi^{real}= e^{\frac{q_R}{2}(|x|^2-|y|^2)-q_S (x \bar y +y\bar x)} \,
\varphi\,.
\ee
In principle $\xi$ and $h$ should be determined asking for the fluctuations to solve the D-term, however it is convenient to define
$U= e^{-fP/2}\varphi$ and express the D-term equations in terms of this new doublet which is related to $\xi$ as
\be\label{defxi}
\xi = i \Psi^{-1} (U-h)\,.
\ee
In the D-term equations the dependence on $h$ will completely drop and it will be fixed asking for regularity of $\xi$ once the solution for $U$ will be known. Here we expand the equations (\ref{uterm}) for $U$ in terms of the
doublets:
\begin{subequations}
\label{ueqsf}
\begin{align}
\label{ueqpf}
\p_x \p_{\bar x} U^+ + \p_y \p_{\bar y} U^+ &{}
-\frac{\mu^2}{D_0} \left(\mu^2 y \p_{\bar y} U^+ + m  \p_{\bar y} U^- \right)
+ \frac{\p_x f}{D_0} \left[(\mu^4 y^2 + m^3x) \p_{\bar x} U^+ + 2m \mu^2 y \p_{\bar x} U^- \right] \nonumber \\
+ q_x \p_{\bar x} U^+ + q_y \p_{\bar y} U^+
& -\left(\mu^4 y\bar y + m^2e^{2f}\right)U^+ + \mu^2 m\left(\bar y + m y \bar x e^{-2f}\right)U^-  =  0\,,\\[2mm]
\label{ueqmf}
\p_x \p_{\bar x} U^- + \p_y \p_{\bar y} U^-  & +
\frac{1}{D_0} \left(m^2\mu^2 y \p_{\bar x} U^+ + m^3 \p_{\bar x} U^- - m^2\mu^2 x \p_{\bar y} U^+ -
\mu^4 y\p_{\bar y} U^- \right) \nonumber \\
+ q_x \p_{\bar x} U^- + q_y \p_{\bar y} U^-
& - \frac{\p_x f}{D_0} \left[(\mu^4 y^2 + m^3x) \p_{\bar x} U^- + 2m^2 \mu^2 xy \p_{\bar x} U^+ \right]  \\
& + \mu^2 m\left(m x\bar y + y e^{2f}\right)U^+ - \left(\mu^4y\bar y + m^4 x\bar x e^{-2f}\right)U^-  =  0\,, \nonumber
\end{align}
\end{subequations}
where in the previous equations we defined $D_0 = \mu^4 (x-y)^2-m^3 x$, $q_x = q_R \bar x - q_S \bar y $ and $q_y = -(q_R \bar y + q_S \bar x)$. To solve these equations which look extremely complicated we need 
some simplifications. First of all since we know the behavior of the Painlev\'e transcendent only in a neighborhood of the origin we will restrict our attention to the region $mr \ll 1$. Second we will look to a particular region in the 
parameter space where $\mu \ll m$. The equations for the doublet $U$
greatly simplify and take the following form
\begin{subequations}
\label{eqmuzerof}
\begin{align}
\label{eqmu0pf}
\p_x \p_{\bar x} U^+ + \p_y \p_{\bar y} U^+ + q_x \p_{\bar x} U^+ + q_y \p_{\bar y} U^+
- \p_x f \p_{\bar x}U^+ - m^2e^{2f} U^+ & = 0\,, \\[2mm]
\label{eqmu0mf}
\p_x \p_{\bar x} U^- + \p_y \p_{\bar y} U^- -
\frac1{x}\, \p_{\bar x} U^-  + \p_x f \p_{\bar x}U^- + q_x \p_{\bar x} U^- + q_y \p_{\bar y} U^-
- m^4 e^{-2f} x\bar x U^- & = 0\,.
\end{align}
\end{subequations}
Since the equations for $U^+$ and $U^-$ are now independent we will solve them separately. We start with $U^+$ which will not admit a localized solution and then move to $U^-$.

\subsubsection*{Solution for $U^+$}
Using the known asymptotic form of the Painlev\'e transcendent in a neighborhood of the origin we find that the equation for $U^+$ becomes
\be
\p_x \p_{\bar x} U^+ + \p_y \p_{\bar y} U^+ +( q_x-  m^2 c^2 \bar x) \p_{\bar x} U^+ + q_y \p_{\bar y} U^+
 - m^2 c^2(1+2m^2 c^2 x \bar x)U^+  = 0\,.
\ee
This equation is in general very difficult to solve, but some simplifications occur if we take $q_S=0$.\footnote{According to the results in appendix B the flux $q_S$ does not affect the convergence of the wavefunction in the $\mathbf{10}$ sector for $\mu\ll m$ so the conclusion that we arrive at should be valid also for $q_S\neq 0$.} In this case we can take the function $U^+$ to be a function of $r = \sqrt{x \bar x}$ times a holomorphic function of $y$
\be
U^+ = g(y) G(r)\,.
\ee
Using this form we see that the equation for $U^+$ becomes an equation for $G(r)$
\be\label{U0plus}
G''(r)+  \frac{1}{r}G'(r)+2r (q_R-c^2 m^2)G'(r)-4c^2 m^2 \left(2 c^2 m^2 r^2+1\right) G(r)=0\,.
\ee
Eq.(\ref{U0plus}) has a regular singular point at $r=0$ and it can be shown easily that at
this point there is an analytic solution and a solution with a logarithmic singularity
that diverges and must be discarded. Up to normalization the analytic solution has
a series expansion
\be
G(r) = 1 + m^2 c^2 r^2 + \dots.
\ee
This function is not localized at $r=0$. Thus, in the following we set $U^+=0$.


\subsubsection*{Solution for $U^-$}
The equations for $U^-$ once we take into account the asymptotics of the Painlev\'e transcendent has the following form
\be
\p_x \p_{\bar x} U^- + \p_y \p_{\bar y} U^- -
\frac1{x}\, \p_{\bar x} U^-  +  (q_x + m^2 c^2 \bar x)\p_{\bar x} U^- + q_y \p_{\bar y} U^- - m^4 c^{-2} x\bar x U^-  = 0\,.
\ee
The solution to this equation is quite simple
\be\label{Usol}
U^- = e^{\lam_{\mathbf{10}} x (\bar x- \zeta_{\mathbf{10}}\bar y)} g_j(y+\zeta_{\mathbf{10}} x)\,,
\ee
where $g_j(y+\zeta_{\mathbf{10}} x)$ are holomorphic family functions, $\lam_{\mathbf{10}}$ is the lowest root of the polynomial
\be
m^4 (\lambda_{\mathbf{10}} -q_R)+\lambda c^2 \left( c^2 m^2 (q_R-\lambda_{\mathbf{10}} )-\lambda_{\mathbf{10}} ^2+q_R^2+q_S^2\right)=0
\ee
and 
\be
\zeta_{\mathbf{10}} = -\frac{q_S}{(\lam_{\mathbf{10}}-q_R)}\,.
\ee
Using (\ref{a1}), (\ref{a2}) and (\ref{defxi}) one gets the physical fluctuations in the main text (\ref{phys10}) which we repeat here for convenience,

\be
\label{phys10A} 
\vec{\vphi}_{\mathbf{10}^+}^j \, =\, \gamma_{\mathbf{10}}^j
\vec v_+
 e^{f/2} \chi_{\mathbf{10}}^j 
 \qquad \quad 
\vec{\vphi}_{\mathbf{10}^-}^j \, =\, \gamma_{\mathbf{10}}^j
\vec v_-
e^{-f/2} \chi_{\mathbf{10}}^j 
\ee
with
\be
\vec{v}_{\mathbf{10}^+}=\left (\begin{array}{c}
\frac{i\lam_{\mathbf{10}}}{m^2}\\
-\frac{i\lam_{\mathbf{10}}\zeta_{\mathbf{10}}}{m^2} \\
0 \end{array}\right )\qquad \vec{v}_{\mathbf{10}^-}=\left (\begin{array}{c}
0 \\
0 \\
1 \end{array}\right ).
\label{vec10}
\ee
The scalar wavefunctions $\chi_{\mathbf{10}}$ read
\be
\chi_{\mathbf{10}}^j \, =\,e^{\frac{q_R}{2}(|x|^2-|y|^2)-q_S (x \bar y +y\bar x)+\lam_{\mathbf{10}} x (\bar x - \zeta_{\mathbf{10}} \bar y)} \, g_j (y +\zeta_{\mathbf{10}}x)
\ee
where $g_j\, =\, m_*^{3-j}(y +\zeta_{\mathbf{10}}x)^{3-j}$ for $i=1,2,3$.
In order to compute the norm for the $\mathbf{10}$ sector we have to include the scalar product in the $\mathfrak{su(2)}$ algebra. Thus, to have canonical kinetic terms we must impose the following matrix is the identity
\bea
K^{ij}_{\mathbf{10}} & = & m_*^2\int_{S}\tr(\vec{\vphi}_{\mathbf{10}^+}^i{}^\dagger\vec{\vphi}_{\mathbf{10}^+}^j+\vec{\vphi}_{\mathbf{10}^-}^i{}^\dagger\vec{\vphi}_{\mathbf{10}^-}^j)d\text{vol}_{S} \\ \nonumber
& = & m_*^2 (\g_{\mathbf{10}}^i)^*\g_{\mathbf{10}}^j \sum_{\kappa = \pm} ||\vec{v}_{\mathbf{10}^\kappa}|| \, \int_S e^{\kappa f} (\chi_{\mathbf{10}}^i)^* \chi_{\mathbf{10}}^j\, d{\text{vol}}_S
\label{4dnorm10A}
\eea
The only non-vanishing elements are those in the diagonal which can be computed using the formulas in appendix A in \cite{fimr12} leading to
\be
|\gamma_{\mathbf{10}}^j|^2=-\frac{c}{m_*^2\pi^2(3-j)!}\frac{1}{\frac{1}{2\lam_{\mathbf{10}}+q_R(1+\zeta_{\mathbf{10}}^2)-m^2 c^2}+\frac{c^2\lam_{\mathbf{10}}^2}{m^4}\frac{1}{2\lam_{\mathbf{10}}+q_R(1+\zeta_{\mathbf{10}}^2)+m^2 c^2}}\left (\frac{q_R}{m_*^2}\right )^{4-j}.
\ee

When computing the wavefunctions and normalization factors we assumed that we could approximate the Painlev\'e function $f$ to second order in the coordinates. Now that we have the wavefunctions we revisit this assumption to see whether this is indeed a consistent approximation. Recall from (\ref{Painori}) the form of the Painlev\'e solution including the neglected higher order term
\be
f(r)=\log c +c^2m^2r^2+m^4r^4\left ( \frac{c^2}{2}-\frac{1}{4c^2}\right )+\dots.
\ee
The additional contribution will become important at $r=r_0$ where it is of the same order or magnitude as the $r^2$ term, namely

\be\label{condP}
c^2m^2r_0^2\sim m^4r_0^4\left | \frac{c^2}{2}-\frac{1}{4c^2}\right |\quad\Longrightarrow\quad r_0^2\sim\frac{4c^4}{|2c^4-1|\,m^2}.
\ee
A consistent approximation requires that at such $r_0$ the wavefunction is sufficiently damped so that the effect of higher order terms is in fact negligible. The damping of the wavefunction can be read off from (\ref{phys10A}) and is controlled by the $|x|^2$ coefficient in the exponential. Thus, the wavefunction will be exponentially small at $r=r_0$ when
\be\label{condP2}
-\left ( \frac{q_R}{2}+\lam_{\mathbf{10}}\pm \frac{m^2c^2}{2} \right )r_0^2\gg 1
\ee
where the term $\pm m^2c^2/2$ comes from the factor $e^{\pm f/2}$ in (\ref{phys10A}). Using the value of the critical radius $r_0$ in (\ref{condP}) we get a condition on the parameters $m$, $q_R$ and $q_S$, namely
\be\label{condP3}
-\left ( \frac{q_R}{2}+\lam_{\mathbf{10}}+ \frac{m^2c^2}{2} \right )\frac{4c^4}{|2c^4-1|\,m^2} \gg 1
\ee
where we took the plus sign in (\ref{condP2}) since it is more restrictive. It is perhaps useful to consider the particular case $q_S=0$ since the above condition becomes more transparent and according to the discussion in appendix \ref{ap:chiral} this should not affect the convergence of the wavefunction. Thus, taking $q_S=0$ one can get a simple solution for $\lam_{\mathbf{10}}$ and (\ref{condP3}) becomes
%
\be\label{condP4}
\sqrt{\left ( \frac{q_R}{m^2}+c^2 \right )^2+\frac{4}{c^2}}\,\frac{2c^4}{|2c^4-1|}  \gg 1.
\ee

\subsection*{Non-perturbative corrections}

In this section we present some details on the computation of the correction to the background and fluctuations in a real gauge due to the non-perturbative effects.

\subsubsection*{Background}

As explained in the main text one can compute the correction to the background using the same strategy as for the tree level solution. Namely, one solves first the F-terms and performs an arbitrary complexified gauge transformation (see eq.(\ref{cgtnp})) on the solution which is then plugged in the D-term equation. This yields an equation for such transformation which in our particular case is given in terms of the functions $f$ and $k$. 

Taking the background fields in a real gauge (\ref{bkgrealeps}) and imposing they satisfy the D-term equation we arrive at the Painlev\'e equation (\ref{eq:ode}) for $f$ and to the following equation for $k$,
\begin{eqnarray}
\nonumber
\cosh f\, \p_x\p_{\bar x}k+\p_xf\p_{\bar x}k\,e^{f}-\p_xk\p_{\bar x}f\,e^{-f}+me^{f}(m^2\bar xk^*-mk+2\bar\th_y\p_{\bar x}f)\\\label{pain2}
-m^2\bar xe^{-f}(m^2xk-mk^*+2\th_y\p_xf)=0.
\end{eqnarray}
One can solve this equation near the origin following the same reasoning given around eq.(\ref{Painori}) according our local approach. Thus, it is enough to know the solution to the Painlev\'e equation for $f$ to second order, namely $f=\log c+m^2c^2x\bar x+\dots$, which yields the solution to $k$ given in the main text,
\be
k=\bar\th_y\, c^2mx+\th_y\,\frac{1-c^2}{2c^2-1}m^2 \bar x^2+\dots 
\ee

\subsubsection*{Sector 5}

This sector in not charged under the $SU(2)\subset E_6$ where the T-brane lives so the analysis of the correction to the physical wavefunction reduces to that of section 5.1 in \cite{fimr12}. The equations of motion read
\bea
\bar\partial_{\langle A\rangle}a&=&0\\
\bar\partial_{\langle A\rangle}\varphi+i[\langle \Phi\rangle,a]+\eps\,\partial\th_0\,\w\,\p_{\langle A\rangle}a&=&0\\
\omega\,\w\,\p_{\langle A\rangle}a-\frac{1}{2}[\langle \bar\Phi\rangle,\varphi]&=&0
\eea
which in a holomorphic gauge reduce to
\bea
\p_{\bar{x}} a_{\bar{y}} - \p_{\bar{y}} a_{\bar{x}} & = & 0 \label{ft03}\\
\p_{\bar{m}} \vphi_{xy} + i 2\mu^2 (x-y) a_{\bar{m}} & = & i\eps  \left[\th_{y} \p_x a_{\bar{m}} -  \th_{x} \p_y a_{\bar{m}}\right] + \CO(\eps^2)
\eea
for the F-terms while the D-term is
\bea
& & \hspace*{-1cm}\left\{\partial_x + \bar xq_R  - \bar y q_S\right\} a_{\bar x}  +  \left\{\partial_y  - \bar yq_R - \bar x q_S \right\} a_{\bar y} - 2i \mu^2 (\bar x-\bar y) \varphi_{xy} \qquad 
 \\
&  = & 
i \eps \bar\th_{x}\left\{y q_R+ xq_S \right\} \varphi_{xy}
-i \eps \bar\th_{y}\left\{- q_Rx + y q_S \right\} \varphi_{xy}.
\nonumber
\eea
The correction to the wavefunction is found to be 
\be 
\varphi_{\mathbf 5}^{(1)} = m_*\g_{\mathbf{5}} e^{(x-y)(\zeta_{\mathbf{5}}\bar x-(\lam_{\mathbf{5}}-\zeta_{\mathbf{5}})\bar y))}\Upsilon_{\mathbf{5}}
\label{chic11}
\ee
with $\lam_{\mathbf{5}}$, $\zeta_{\mathbf{5}}$ defined as in (\ref{wave5p}) which depend on the subsector $\mathbf{5}_r$, $r=1,2$, and
\be
\Upsilon_{\mathbf{5}}=-\frac{1}{4\mu^2}(\zeta_{\mathbf{5}} \bar{x} - (\lam_{\mathbf{5}} - \zeta_{\mathbf{5}})\bar{y})^2(\th_x+\th_y)+\frac{\delta_1}{2}(x-y)^2+\frac{\delta_2}{\zeta_{\mathbf{5}}}(x-y)(\zeta_{\mathbf{5}} y+(\lam_{\mathbf{5}}-\zeta_{\mathbf{5}})x)
\ee
with the constants $\delta_1$ and $\delta_2$ given by 
\bea\label{d1}
\delta_1&=& \frac{2\mu^2}{\lam_{\mathbf{5}}^2}\{ \bar \th_x(q_R(\zeta_{\mathbf{5}}-\lam_{\mathbf{5}}) + q_S\zeta_{\mathbf{5}} ) +\bar\th_y(q_R\zeta_{\mathbf{5}} - q_S(\zeta_{\mathbf{5}}-\lam_{\mathbf{5}}))\}\\
\delta_2&=& \frac{2\mu^2\zeta_{\mathbf{5}}}{\lam_{\mathbf{5}}^2}\{ \bar \th_x(q_R+q_S) +\bar\th_y(q_R-q_S)\}.
\label{d2}
\eea
The holomorphic terms in $\Upsilon_{{\mathbf{5}}}$, which depend on $\bar\th_{x}$ and $\bar\th_{y}$ through $\d_1$ and $\d_2$,
are needed to satisfy the corrected D-term equation. Going back to a real gauge we arrive at eq.(\ref{wave5np}) in the main text.

\subsubsection*{Sector 10}

As argued in the main text there is no actual need to compute the corrections to the $\mathbf{10}$ sector in the $\mu\ll m$ approximation since these do not modify the normalization factors or induce mixing. Here we complete the argument by showing that in eq.(\ref{phi10np}) $U^{(1)}_-=0$.

We know from the discussion below eq.(\ref{U0plus}) that $U^{(0)}_+=0$, which implies
$\xi^{(0)}_-=0$. Then, from (\ref{solxinp}) it follows
\be
\xi^{(1)}_+ = \frac{i}{m^2 x} U^{(1)}_-.
\label{xi1p}
\ee
Moreover, from (\ref{cgtnp}) and (\ref{Fsolnp}) we obtain
\be
a^{(1)}_+ = \frac{i}{m^2 x} \, e^{f/2} \, \bar \p U^1_- \qquad ; \qquad
\varphi^{(1)}_- = e^{-f/2} \, U^{(1)}_-
\label{a1phi1}
\ee
which when substituting in the D-term yield
\be
\p_x \p_{\bar x} U^{(1)}_- -
\frac1{x}\, \p_{\bar x} U^{(1)}_-  + \p_x f \p_{\bar x}U^{(1)}_- +q_x\p_xU^{(1)}_-+q_y\p_yU^{(1)}_-- m^4 e^{-2f} x\bar x U^{(1)}_- = 0.
\label{dtuminus}
\ee
This equation is the same as (\ref{eqmu0mf}) so the solution near the origin is given by $U^{(1)}_-=e^{\lam_{\mathbf{10}}x(\bar x-\zeta_{\mathbf{10}}\bar y)}s(y)$ which is localized. However, in order to ensure that $\xi^{(1)}_+$ given in (\ref{xi1p}) is regular at $x=0$ it must be that $s \equiv 0$. Thus, we take $U^{(1)}_-=0$ which yields the particular form (\ref{phys10np}) for the correction.


\section{Doublet-triplet splitting and local chirality}
\label{ap:chiral}

When constructing the $E_6$ model in the main text we made some particular choices of fluxes to have the correct chiral modes and doublet-triplet splitting. In this appendix we discuss in more detail these choices. 

The Higgs background $\langle\Phi  \rangle$ generally produces matter curves $\Sigma_{\rho}$ where we find localized modes in the transverse direction for both chiralities. Only when non-trivial gauge fluxes are included a specific chirality is selected. More precisely, to have a chiral spectrum in the sector $\rho$ living at $\Sigma_\rho$ we must ensure that 
\be\label{Gchi}
\int_{\Sigma_\rho}\tr\,\langle F_{\rho} \rangle\neq0.
\ee
%
This condition requires a global knowledge of the matter curve $\Sigma_\rho$ along $S_{GUT}$ and the flux along it. However, we cannot impose (\ref{Gchi}) in practice since we do not know the geometry or fluxes away from the Yukawa point so we need an alternative characterization of chirality suited to our local approach. The notion of local chirality was discussed in \cite{palti12} and it boils down to demanding that matter wavefunctions of a certain 4d chirality are localized near the Yukawa point $p$. Indeed, when gauge fluxes are included such that the wavefunction for a given sector $\rho$ is localized in the region around $p$, its conjugate sector $\bar\rho$ will not contain any localized mode in that same region, and this signals a net local  chirality. 


A simple way to obtain a condition on the fluxes to have local chirality is to first T-dualize in the $z,\,\bar z$ directions to a system of magnetized D9 branes and look at the index theorem in 6d. As explained in appendix A of \cite{afim} under T-duality we get a gauge field along $\bar z$, $\langle A_{\bar z} \rangle=\langle \Phi_{xy}\rangle$, so we end up with magnetic fluxes $F_{x\bar z}=D_x\Phi_{xy}$ and $F_{y\bar z}=D_y\Phi_{xy}$. For T-brane backgrounds we also have a flux along the $z\bar z$ direction, $F_{z\bar z}=i[\Phi_{xy},\Phi_{xy}^\dagger]$.

The Dirac index for a given representation $\mathcal R$ in 6d reads 
\be\label{Dind}
{\rm index}_{\mathcal R}  \slashed D=\frac{1}{48(2\pi)^2}\int \left ( \tr_{\mathcal R}\,F\,\w\, F\,\w \,F-\frac{1}{8}\tr_{\mathcal R}\,F\,\w\, \tr R\,\w\, R \right )
\ee
where $F$ is the gauge flux and $R$ is the Riemann tensor. The local chirality notion in this setup translates into asking that the integrand is different from zero at the Yukawa point. Since the second term in (\ref{Dind}) vanishes for flat spaces we should look at the quantity\footnote{We take $F=F_{\a\bar\b}\,dx^\a\,\w\, d\bar x^{\bar \b}$ so we include a factor of $i$ to make $\tr_{\mathcal R} F^3_{x\bar xy\bar yz\bar z}$ a real number.}
\begin{eqnarray}\label{indR}
\mathcal I_{\mathcal R}\equiv\frac{i}{6}\tr_{\mathcal R}\,\left (F\,\w\, F\,\w \,F\right )_{x\bar xy\bar yz\bar z}=i\,\tr_{\mathcal R}\big ( F_{x\bar x}\{F_{y\bar y},F_{z\bar z}\}+F_{x\bar z}\{F_{y\bar x},F_{z\bar y}\}+ \\\nonumber
F_{x\bar y}\{F_{y\bar z},F_{z\bar x}\}- \{F_{x\bar x},F_{y\bar z}\}F_{z\bar y} - \{F_{x\bar y},F_{y\bar x}\}F_{z\bar z} - \{F_{x\bar z},F_{y\bar y}\}F_{z\bar x}  \big ).
\end{eqnarray}
The condition to have local chirality in a given sector $\rho$ is that $\mathcal I_{\mathcal R}<0$ for such sector in a given region.\footnote{The label $\rho$ refers to the gauge numbers with respect to the unbroken gauge group while $\mathcal R$ is the representation under the broken part. Thus, $\mathcal I_{\mathcal R}$ depends on both $\rho$ and $\mathcal R$.} For the case of intersecting branes the flux $F_{z\bar z}$ vanishes and the representation $\mathcal R$ is abelian so $\mathcal I_{\mathcal R}$ reduces to the criterion for local chirality discussed in \cite{palti12} and \cite{fimr12}. 

In the following we compute $\mathcal I_{\mathcal R}$ for every sector in the $E_6$ model which gives the conditions on the fluxes to have both chirality and doublet-triplet splitting.

\subsubsection*{Sector 10}

For the $\mathbf{10}$ sector the representation $\mathcal R$ under the broken part $\mathfrak{su}(2)\oplus\mathfrak{u}(1)$ is $\mathbf {2}_{-1}$ so the fluxes are $2\times 2$ matrices which we write in terms of $Q,\,P,\,E^+,\,E^-$ defined by
\be\nonumber
Q=\left (\begin{array}{cc}
1&0\\
0&1\end{array}\right )\qquad P=\left (\begin{array}{cc}
1&0\\
0&-1\end{array}\right )\qquad
E^+=\left (\begin{array}{cc}
0&1\\
0&0\end{array}\right )\qquad E^-=\left (\begin{array}{cc}
0&0\\
1&0\end{array}\right ).
\ee
Taking into account the T-brane background (\ref{Tbranebkg}) together with the primitive fluxes (\ref{sumFp}) we have
\be\nonumber
\begin{array}{ll}
-iF_{x\bar x}=-q_R(\mathbf{10}_i) Q-m^2c^2P&-iF_{x\bar y}=q_S(\mathbf{10}_i) Q\\
-iF_{y\bar y}=q_R(\mathbf{10}_i)Q&-iF_{x\bar z}=ib\mu^2 Q-2im^2c^2\bar x(mcE^+-\frac{m^2x}{c}E^-)  \\
-iF_{z\bar z}=m^2c^2P&-iF_{y\bar z}=-i\mu^2Q
\end{array}
\ee
where the index $i$ runs through all the particles with different hypercharge within the $\mathbf{10}$ multiplet according to table {\ref{t:sectors}} in the main text. Using the expression in (\ref{indR}) we find that for this sector the local index is
\be
\mathcal I_{\mathbf {2}_{-1}}= -2m^4c^4q_R(\mathbf{10}_i)-4b\mu^4q_S(\mathbf{10}_i)+2(1-b^2)\mu^4q_R(\mathbf{10}_i)
\ee
where we dropped some $x$-dependent terms since these are negligible compared to the constant terms near the origin. In fact, when computing localized wavefunctions in Section \ref{s:zeromodes} we take the approximation $\mu\ll m$ which yields
\be
\mathcal I_{\mathbf {2}_{-1}}\approx -2m^4c^4q_R(\mathbf{10}_i).
\ee
Thus, the condition $\mathcal I_{\mathbf {2}_{-1}}<0$ translates into $q_R(\mathbf{10}_i)>0$ for every $i$. Taking into account the different values of the hypercharge we arrive at eq.(\ref{chiral10}) in the main text. Notice that this condition can also be read off from the physical wavefunction (\ref{phys10}) which is only convergent for $q_R>0$.

\subsubsection*{Sector 5 and doublet-triplet splitting}

The $\mathbf 5$ sector contains the Higgs doublet and a triplet that should not be present at low energies. This means that we must include a hypercharge flux such that it yields a chiral spectrum for the doublets while keeping the triplet sector non-chiral. In terms of local chirality, the index (\ref{indR}) should be negative for the doublets and vanishing for the triplets. Since this sector transforms as $\mathbf 1_2$ under the broken $\mathfrak{su}(2)\oplus\mathfrak u(1)$ it does not feel the T-brane background and the index reduces to the expression in \cite{fimr12}, namely
\be
\mathcal I_{\mathbf 1_2}=\det \mathbf m_{\mathbf 5_i}=\det\left (\begin{array}{ccc}
F_{x\bar x}&F_{y\bar x}&F_{z\bar x}\\
F_{x\bar y}&F_{y\bar y}&F_{z\bar y}\\
F_{x\bar z}&F_{y\bar z}&0 \end{array}\right ).
\ee
Given the Higgs background (\ref{Tbranebkg}) and fluxes (\ref{sumFp}) the matrix $\mathbf m_{\mathbf 5_i}$ is  
\be
\mathbf m_{\mathbf 5_i}=\left (\begin{array}{ccc}
-q_R(\mathbf 5_i)&q_S(\mathbf 5_i)&2ib\mu^2\\
q_S(\mathbf 5_i)&q_R(\mathbf 5_i)&-2i\mu^2\\
-2ib\mu^2&2i\mu^2&0 \end{array}\right )
\ee
which yields
\be
\mathcal I_{\mathbf 1_2}=-8\mu^4b\,q_S(\mathbf 5_i)+4\mu^4(1-b^2)q_R(\mathbf 5_i).
\ee
For the particular choice $b=1$ the second term vanishes and $\mathcal I_{\mathbf 1_2}=-8\mu^4q_S(\mathbf 5_i)$ so in order to have doublet triplet splitting we must ensure
\be
q_S(\mathbf 5_1)>0,\quad q_S(\mathbf 5_2)=0\quad\Longrightarrow\quad N_Y+6N=0,\quad N>0.
\ee
Again, one can check that this condition is precisely what is needed to make the physical wavefunction (\ref{wave5p}) convergent. On the other hand, for the triplets the wavefunction is localized only in the transverse direction to the matter curve but not in the longitudinal one.

Notice that the choice $b=0$ is troublesome since it does not allow to have doublet-triplet splitting and chiral quarks. Indeed, for $b=0$ the condition for having non-chiral triplets is $q_R(\mathbf 5_2)=-2M+\frac{1}{3}\tilde N_Y=0$. However, this is incompatible with having chirality in the $\mathbf{10}_2$ subsector which requires $q_R(\mathbf{10}_2)=M-\frac{1}{6}\tilde N_Y>0$.


\section{Elliptic fibration for the $E_6$ singularity}
\label{ap:ellip}

In the main text a local description of the GUT divisor has been used without any reference to its embedding into a three-fold used for the compactification. In this appendix using deformation of ADE singularities we will be 
able to have a local description of the geometry of the elliptic fibration around the $E_6$ point and have a further check of the location of the matter curves. We start recalling that the general form of an unfolded $E_6$
singularity is
\be\label{ell}
Y^2 = X^3 + X( \eps_2 z^2 +\eps_5 z + \eps_8)+ \left( \frac{z^4}{4}+\eps_6 z^2 +\eps_9 z + \eps_{12}\right)\,.
\ee
Here $X,Y \in \mathbb{C}^2$ are coordinates in the elliptic fiber\footnote{Here we are describing the elliptic curve in an affine patch so that $X$ and $Y$ are inhomogeneous coordinates. However it is easy to go to the usual
Weierstra$\ss$ form of the elliptic fiber taking the projective closure of  (\ref{ell}) in $\mathbb{P}_{1,2,3}$. If we call the homogenous coordinates of $\mathbb{P}_{1,2,3}$ $(u,v,w)$ then we have $X=v u^{-2}$ and $Y=w u^{-3}$
in the affine patch $\mathbb{P}_{1,2,3}\smallsetminus \mathcal{Z}(u).$} and $z$ is a local coordinate in the base manifold. The Casimir invariants of $E_6$ whose explicit expression can be found in the appendices of \cite{Katz}, will 
be determined by a particular choice of Higgs background on the GUT divisor. It is convenient to define
\be
f= \eps_2 z^2 +\eps_5 z + \eps_8\,, \quad g= \frac{z^4}{4}+\eps_6 z^2 +\eps_9 z + \eps_{12} \,.
\ee
Now inspecting the equation defining the elliptic fiber we can see that it will be singular whenever
\be
\Delta =  27 g^2+4 f^3 =0\,.
\ee
If we specialize to the Higgs background presented in the main text we find that the Casimir invariants have the following expression
\begin{subequations}
\begin{align}
\eps_2=&\frac{1}{6} \left(-3 m^3 x-5 \mu ^4 (x-y)^2\right)\\
\eps_5=&-\frac{8}{81} \mu ^6 (x-y)^3 \left(15 m^3 x+\mu ^4 (x-y)^2\right)\\
\eps_6=&\frac{1}{1944}\big[81 m^9 x^3-135 \mu ^4 m^6 x^2 (x-y)^2-1125 \mu ^8 m^3 
  x (x-y)^4+155 \mu ^{12} (x-y)^6\big]\\
\eps_8=&\frac{1}{34992}\big[-729 m^{12} x^4+4860 \mu ^4 m^9 x^3 (x-y)^2-15390 \mu ^8 m^6 x^2 (x-y)^4-\nonumber\\&5460 \mu ^{12} m^3 x (x-y)^6+335 \mu ^{16} (x-y)^8\big]\\
\eps_9=&\frac{2 \mu ^6 (x-y)^3}{19683} \big(1215 m^9 x^3-4941 \mu ^4 m^6 x^2 (x-y)^2-\nonumber\\&675 \mu ^8 m^3 x (x-y)^4+305 \mu ^{12} (x-y)^6\big)\\
\eps_{12}&=\frac{1}{5668704}\big[6561 m^{18} x^6-65610 \mu ^4 m^{15} x^5 (x-y)^2+\nonumber\\&317115 \mu ^8 m^{12} x^4 (x-y)^4-536220 \mu ^{12} m^9 x^3 (x-y)^6-289305 \mu ^{16} m^6 x^2 (x-y)^8+
\nonumber\\&27846
   \mu ^{20} m^3 x (x-y)^{10}+15325 \mu ^{24} (x-y)^{12}\big]
\end{align}
\end{subequations}

In order to analyze the singularity it is convenient to define a shifted variable $z'= z-\frac{1}{27} \mu ^2 (x-y) \left(9 m^3 x+7 \mu ^4 (x-y)^2\right)$. In terms of $z'$ the discriminant takes the form:
\be
\Delta = -\frac{1}{8} z'^5 \left[\mu ^2 (x-y) \left(m^3 x-\mu ^4 (x-y)^2\right)^4\right]+\mathcal{O}\left(z'^6\right)\,.
\ee
Thus we can conclude that the fiber will be singular at $z'=0$, and moreover the singularity will enhance at the loci $x=y$ and $m^3x=\mu^4 (x-y)^2$. It is quite simple to analyze these singularities using Kodaira classification
\begin{center}
\begin{tabular}{|c| c| c|c|c|}
\hline
 & ord($f$) & ord($g$) & ord($\Delta$)&Singularity\\
 \hline
 \hline
 $z'=0$ & 0 & 0 & 5 & $A_4$\\
 \hline
 $\left.\begin{array}{c} z'=0\\ x=y\end{array}\right.$ & 0 & 0 & 6 & $A_5$\\
\hline
 $\left.\begin{array}{c} z'=0\\m^3x=\mu^4 (x-y)^2\end{array}\right.$ & 2 & 3 & 7 & $D_5$\\
 \hline
 $\left.\begin{array}{c} z'=0\\ x=y=0\end{array}\right.$ & $\infty$ & 4 & 8 & $E_6$\\
 \hline
\end{tabular}
\end{center}
As a further check of the structure of the fiber over the discriminant locus we can try to resolve the singularity and analyze the intersection pattern of its components. It is first convenient to pass from the Weierstra$\ss$ form to the Tate form of the fibration. This can be achieved using the following change of variables
\be
(X,Y) \rightarrow \left(X+\frac{1}{12} \left(m^3 x-\mu ^4 (x-y)^2\right)^2-\frac{2}{3} \mu ^2 z' (x-y),Y+\frac{1}{2} X \left(m^3 x-\mu ^4 (x-y)^2\right)-\frac{z'^2}{2}\right)\,.
\ee
This change of variables gives the following elliptic fiber:
\be\label{Tate}
Y^2+a_1 XY+a_3 Y= X^3 +a_2 X^2 +a_4 X+a_6\,,
\ee
where:
\be
a_1= m^3 x-\mu ^4 (x-y)^2\,, \quad a_2 =2 \mu ^2 z' (y-x)\,, \quad a_3 = -z'^2\,, \quad a_4 = a_6 =0\,.
\ee
We note that since $a_6=0$ our fibration is a case of the so-called $U(1)$-restricted Tate model which was introduced in \cite{Grimm:2010ez}.\footnote{These kind of models admit a global section in addition to the usual section of the elliptic fibration and this introduces additional massless $U(1)$ generators in the spectrum. 
However this is an artifact of the choice of minimal $E_6$ singularity in (\ref{ell}). If we had added a term $Z^5$ then $a_6$ would no longer be zero.} An explicit resolution of this class of fibrations was given in \cite{Krause:2011xj} using toric methods: of particular interest is the Yukawa point where the extended Dynkin 
diagram of 
$E_6$ does not appear in any possible toric resolution.\footnote{The resolution of singularities in the context of $SU(5)$ models and the appearance of non-Kodaira fibers has also been studied in \cite{Esole:2011sm,Marsano:2011hv,Hayashi:2013lra}. For a systematic analysis of the resolution of singularities of Tate models
and the appearance of exotic fibers see \cite{Lawrie:2012gg}.}\footnote{There are six different resolutions of (\ref{Tate}) that come from different triangulations of the toric ambient space. The actual number of triangulations of the toric ambient space is larger but some triangulations become equivalent once we restrict to the Calabi-Yau hypersurface.} The fact
that we can not recover the extended Dynkin diagram of $E_6$ matches a distinctive feature of T-brane backgrounds, see footnote \ref{e6p}. In fact if the complex structure of the Calabi-Yau hypersurface is tuned to avoid monodromy like in \cite{Braun:2013cb} it is possible to find a resolution of the singularities that
lead to the extended Dynkin diagram of $E_6$.

\end{document}